\documentclass[12pt, journal, onecolumn]{IEEEtran}

\usepackage{times}

\usepackage{amsmath,booktabs}
\usepackage{graphicx}
\usepackage{amsfonts}
\usepackage{algorithmic}
\usepackage{algorithm}
\usepackage{graphicx}
\usepackage{mathptmx}
\usepackage{amsmath}
\usepackage{amssymb}
\usepackage{theorem}
\usepackage{cite}
\usepackage{array}
\usepackage{url}
\usepackage{cite}
\usepackage{subfigure}
\usepackage{epsf,epsfig}

\def\BibTeX{{\rm B\kern-.05em{\sc i\kern-.02
5em b}\kern-.08em\kern-.1667em\lower.7ex\hbox{E}\kern-.125emX}}

\DeclareMathOperator*{\Max}{maximize}
\DeclareMathOperator*{\Min}{minimize}
\usepackage{setspace}
\begin{document}

\title{Benchmarking Practical RRM Algorithms for D2D Communications in LTE Advanced}
%
%
%
%

\author{
\IEEEauthorblockN{Marco Belleschi\IEEEauthorrefmark{1},
G\'{a}bor Fodor\IEEEauthorrefmark{1},
Demia Della Penda\IEEEauthorrefmark{2},
Mikael Johansson\IEEEauthorrefmark{2},
Aidilla Pradini\IEEEauthorrefmark{2},
Andrea Abrardo\IEEEauthorrefmark{3}}\\

\IEEEauthorblockA{\IEEEauthorrefmark{1}
Ericsson AB, Sweden
}\\
\IEEEauthorblockA{\IEEEauthorrefmark{2}
KTH, Royal Institute of Technology, Sweden\\
}
\IEEEauthorblockA{\IEEEauthorrefmark{3}University of Siena, Italy\\
}
}
\markboth{Submitted to IEEE Journal on Selected Areas in Communication}%
{Shell \MakeLowercase{\textit{et al.}}: Bare Demo of IEEEtran.cls
for Journals}

\maketitle
\thispagestyle{empty}
\pagestyle{empty}

\begin{abstract}
Device-to-device (D2D) communication
integrated into cellular networks is a means to
take advantage of the proximity of devices and allow for reusing cellular resources and
thereby to increase the user bitrates and the system capacity.
However, when D2D (in the $3^{rd}$ Generation Partnership Project also called Long Term Evolution (LTE) Direct)
communication in cellular spectrum is supported, there is a need
to revisit and modify the existing radio resource management (RRM) and power control (PC) techniques
to realize the potential of the proximity and reuse gains and to limit the interference
at the cellular layer.
In this paper, we examine the performance of the flexible LTE PC tool box
and benchmark it against a utility optimal iterative scheme.
We find that the open loop PC scheme of LTE performs well for cellular users both in terms of the used
transmit power levels and the achieved signal-to-interference-and-noise-ratio (SINR) distribution.
However, the performance of the D2D users as well as the overall system throughput can be boosted
by the utility optimal scheme, because the utility maximizing scheme takes better advantage of
both the proximity and the reuse gains.
Therefore, in this paper we propose a hybrid PC scheme, in which cellular users employ
the open loop path compensation method of LTE, while D2D users use the utility optimizing
distributed PC scheme.
In order to protect the cellular layer, the hybrid scheme allows for limiting the
interference caused by the D2D layer at the cost of having a small impact on the
performance of the D2D layer. To ensure feasibility, we limit the number of iterations
to a practically feasible level.
We make the point that the hybrid scheme is not only near optimal, but it also allows
for a distributed implementation for the D2D users, while preserving the LTE PC scheme
for the cellular users.
\end{abstract}

\section{Introduction}
\label{Sec:Intro}
Device-to-device (D2D) communication in cellular spectrum supported by a cellular
infrastructure has the potential of increasing spectrum and energy efficiency
as well as allowing new peer-to-peer services by taking advantage of the so called
proximity and reuse gains \cite{Janis:09}, \cite{Doppler1}, \cite{Fodor:12}, \cite{Belleschi:11}.
In fact, D2D (Long Term Evolution (LTE) Direct) communication in cellular spectrum is currently studied by the
$3^{rd}$ Generation Partnership Project (3GPP) to facilitate {\it proximity aware}
internetworking services \cite{Flarion}, national security and public safety
applications \cite{RWS-120003} and machine type communications \cite{M2M}.

\begin{figure}[ht!]
\begin{center}
\includegraphics[width=0.80\hsize]{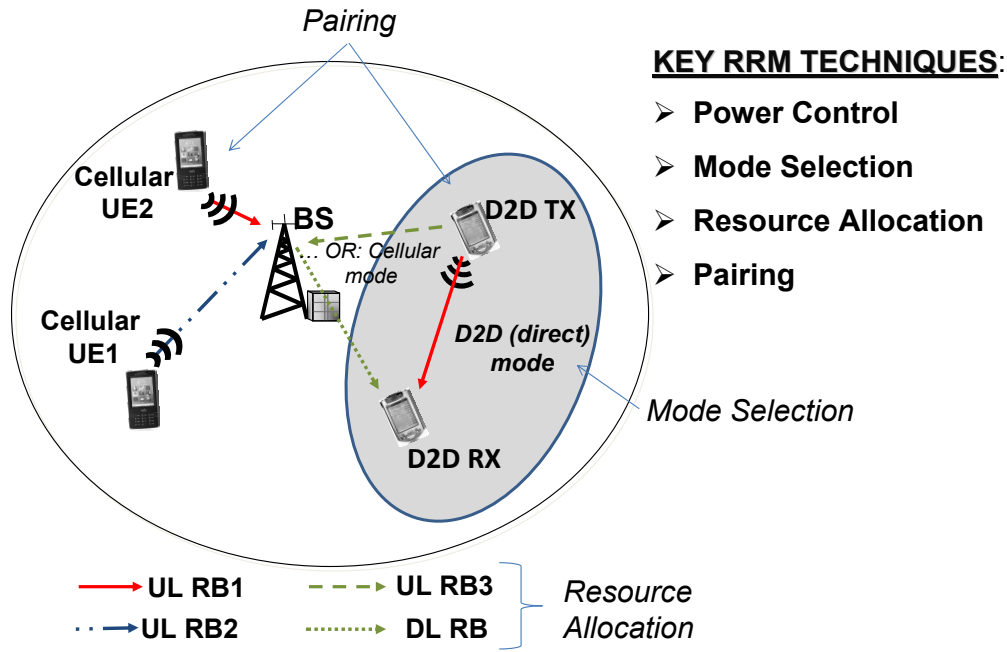}
\caption{A D2D {\it candidate} pair consists of a D2D transmitter and a D2D receiver that
are in the proximity of each other. The mode selection (MS) algorithm needs to decide
on one of 3 possible communication modes:{\it cellular mode},
{\em D2D mode with dedicated resources}, or {\em D2D mode with reused resources}.
This latter case involves a decision on which D2D pair(s) is (are) sharing resources with
which cellular UE (pairing).
}
\label{Fig:ModeSel}
\end{center}
\end{figure}

Obviously, D2D communications utilizing cellular spectrum poses
new challenges, because relative to cellular communication scenarios,
the system needs to cope with new interference situations.
For example, in an orthogonal frequency division multiplexing (OFDM) system
in which user equipments (UE) are allowed to use D2D (LTE direct mode)
communication, D2D communication links may {\it reuse} some of the OFDM
time-frequency physical resource blocks (RB).
Due to the reuse, intracell orthogonality is
lost and intracell interference can become severe due to the random positions of the
D2D transmitters and receivers as well as of the cellular UEs communicating with their
respective serving base stations (BS) \cite{Peng}, \cite{Janis:09-1}.
To realize the promises of D2D communications and to deal with intra- and intercell
interference, the research community has proposed a
number of important radio resource management (RRM) algorithms (see Figure~\ref{Fig:ModeSel}).

Although the objectives of such algorithms may be different (including enhancing the
network capacity \cite{Min-11:2}, improving the reliability \cite{Min-11:1},
minimizing the sum transmission power \cite{Belleschi:11}, ensuring quality of service
\cite{Xiao:11} or protecting the cellular layer (i.e. the cellular UEs)
from harmful interference caused by the
D2D layer \cite{Gu:11}), there seems to be a consensus that the key RRM techniques
include:

\begin{enumerate}
\item
Mode Selection (\textbf{MS}): MS algorithms determine whether D2D candidates in the proximity of each other
should communicate in {\em direct mode} using the D2D link or in {\em cellular mode}
(i.e.\ via the BS) \cite{Doppler2, Hakola:10, Fodor:11}, see Figure \ref{Fig:ModeSel};
\item
Resource Allocation (\textbf{RA}): Surprisingly, resource allocation in the sense of selecting particular OFDM
RBs or frequency channels out of a set of available ones for each transmit-receive pair (cellular or D2D)
is seldom addressed in the literature (\cite{Peng, Wang:11, Yu:11});
\item
\textbf{Pairing}: In the D2D context, pairing refers to selecting the D2D pair(s) and at most one
cellular UE that share (reuse) the same OFDM RB, similarly to multiuser MIMO techniques.
Pairing is a key technique to achieve high reuse gains \cite{Belleschi:11};
\item
Multiple Input Multiple Output (\textbf{MIMO}) Schemes: Interference avoiding MIMO schemes
have been proposed by \cite{Janis}. Such schemes can be applied, for example for the cellular
transmissions to avoid generating interference to a D2D receiver.
\end{enumerate}

Apart from mode selection and resource allocation (i.e.\ RB selection), power control (PC) is a key technique
to deal with intra- and intercell interference
\cite{Xiao:11}, \cite{Gu:11}, \cite{ChiaHao2}, \cite{Xing:10}.
References \cite{Xiao:11} and \cite{ChiaHao2}
analyze the single (isolated) cell scenario and provide some basic insights into the impact
of PC and RA.
The authors of \cite{Gu:11} study a multi-cell
system focusing on a PC scheme that helps minimize the interference from the
D2D layer to the cellular users assuming that D2D users that operate in D2D mode reuse the
cellular resources.
The work reported in \cite{Xing:10} evaluates the LTE PC
scheme for the hybrid cellular and D2D system and concludes that PC needs to
be complemented by mode selection, resource scheduling and link adaptation to properly
handle intra- and intercell interference.

In this paper we examine the performance of the LTE power control scheme when applied to
the hybrid cellular D2D system and compare it with the performance of a distributed power
control scheme based on utility maximization, where dynamic resource allocation and mode selection are
also exercised by the network.
The purpose of this examination is to gain insight into
the applicability of LTE PC for D2D communications by quantifying its
performance with respect to a utility optimal scheme.

We structure the paper as follows.
Section \ref{Sec:LTE} contains a brief overview of the LTE PC toolkit that provides various options for D2D PC.
Section \ref{Sec:Model} describes the system model and states some basic assumptions.
Next, Section \ref{Sec:SINR} elaborates on the signal-to-noise-and-interference-ratio (SINR) target setting and PC problem in the
integrated cellular and D2D environment.
Section \ref{Sec:Decompose} proposes a solution approach to the PC problem based
on the convexification and decomposition of the problem.
Section \ref{Sec:MS} describes the mode selection and resource allocation problem, while
Section \ref{Sec:Heuristic} develops a heuristic aiming at reducing intracell interference
based on full path gain matrix knowledge at the base station,
and two other heuristics that are applicable in real systems.
The numerical results are presented and discussed in Section \ref{Sec:Num}.
Finally, Section \ref{Sec:Conc} concludes the paper.

\section{Power Control Options Based on LTE Mechanisms}
\label{Sec:LTE}
It is natural to base a PC strategy for D2D communications
{\em underlaying}
an LTE network on the LTE standard uplink PC mechanisms~\cite{Doppler1}.
Building on the already standardized and widely deployed schemes facilitates not only a
smooth introduction of D2D-capable
user equipment (UE),
but would also help to develop inter-operable solutions between different
devices and network equipments.
However, due to intracell interference and new intercell
interference scenarios, the question naturally arises whether the available LTE PC is suitable for D2D communications integrated in
an LTE network.
Also, the ad-hoc networking community has proposed efficient distributed schemes
suitable for D2D communications, including situations with or without the availability
of a cellular infrastructure (see e.g. \cite{Soldati:09},
\cite{ChiaHao2}, \cite{Xing:10}, \cite{ChiaHao1}).
Such schemes can also serve as a basis for D2D PC design.

The LTE PC scheme can be seen as a `toolkit' from which different PC strategies
can be selected depending on the deployment scenario and operator preference \cite{Arne}.
It employs a combination of open-loop (OL) and closed-loop (CL) control
to set the UE transmit power (up to a maximum level of $P_{MAX}=24$ dBm) as follows:
\begin{equation}
\label{Eq:LTEPC}
P^{\text{UE}} =\text{min}\Big[P_{MAX}, \underbrace{P_0 - \alpha \cdot G}_\text{OL operating point}
+ \underbrace{\Delta_{\text{TF}} + f\big(\Delta_{\text{TPC}}\big)}_\text{dynamic offset}
+ \underbrace{10  \cdot \text{log}_{10} M}_\text{BW factor}\Big],
\end{equation}

\noindent where $G$ is the path gain between the UE and the BS.
The OL operating point allows for {\it path loss (PL) compensation} and the
dynamic offset can further adjust the transmit power
taking into account the current modulation and coding scheme (MCS)
and explicit transmit power control (TPC) commands from the network.
The bandwidth factor takes into account the number of scheduled RBs ($M$).
For the OL operating point,
$P_0$ is a base power level used to control the SNR target
and it is calculated as \cite{R1-074850}:
\begin{equation}
\label{Po}
P_0=\alpha\cdot(\gamma^{tgt}+P_{IN})+ (1-\alpha)\cdot(P_{MAX} -10  \cdot \text{log}_{10} M ),
\end{equation}
where $\alpha$ is the PL compensation factor
and $P_{IN}$ is the estimated noise and interference power.
For the dynamic offset, $\Delta_{\text{TF}}$ is the transport format
(MCS) dependent component, $f\big(\Delta_{\text{TPC}}\big)$ represents the explicit
TPC commands.

For the integrated D2D communications scenario,
we consider the following options:

\begin{itemize}
\item
No Power Control (NPC), reference case:
With NPC, there is no fixed $\gamma^{tgt}$ and the transmit power of the cellular UEs
and D2D transmitters is set to some fixed value that is equal to or less than $P_{MAX}$
according to \eqref{Eq:LTEPC}\footnote{Note that \eqref{Po} is valid only in the case when
a $\gamma^{tgt}$ value exists.}.
For $M=1$ this can be obtained by setting $\alpha=0$ and $P_{MAX}=P_{fix}$ in \eqref{Po}.
\item
Fixed SNR target (FST): FST fully utilizes the LTE path loss compensation capability
by setting $\alpha=1$ and $P_0=\gamma^{tgt}+P_{IN}$, where $\gamma^{tgt}$ is a predefined
SNR target and $P_{IN}$ is the interference plus noise power (in practice, for simplicity,
$P_{IN}$ can be set to a fixed value, e.g.\ $P_{IN} \approx -121...-116 \text{ dBm}$).
\item
Open Loop with Fractional Path Loss Compensation (OFPC):
The OFPC scheme allows users to transmit with variable power levels, depending on their path loss.
In contrast to the FST-case, the OFPC compensates for the fraction of the path loss by setting
$\alpha$ to some suitable value in the range $[0,1]$, e.g.\ 0.4 \dots 0.9.
\item
Closed Loop PC (CL): CL extends the FST scheme by adding the dynamic offset
or tuning step $f\big(\Delta_{\text{TPC}}\big)$ in (\ref{Eq:LTEPC}) in order to
compensate the measured SINR ($\hat{\gamma}$)
at the receiver with the desired SNR target value.
The tuning step can be computed as follows \cite{Xing:10}:

\begin{equation}
f\big(\Delta_{\text{TPC}}\big)=\left\{
\begin{array}{rl}
\left|\gamma^{tgt}-\hat{\gamma}\right|/2 & \text{if } \left|\gamma^{tgt}-\hat{\gamma}\right| > 2\text{ dB} \\
1 \text{ dB} & \text{ otherwise }
\end{array}
\right.
\end{equation}
\end{itemize}

\begin{figure}[ht!]
\begin{center}
\includegraphics[width=0.80\hsize]{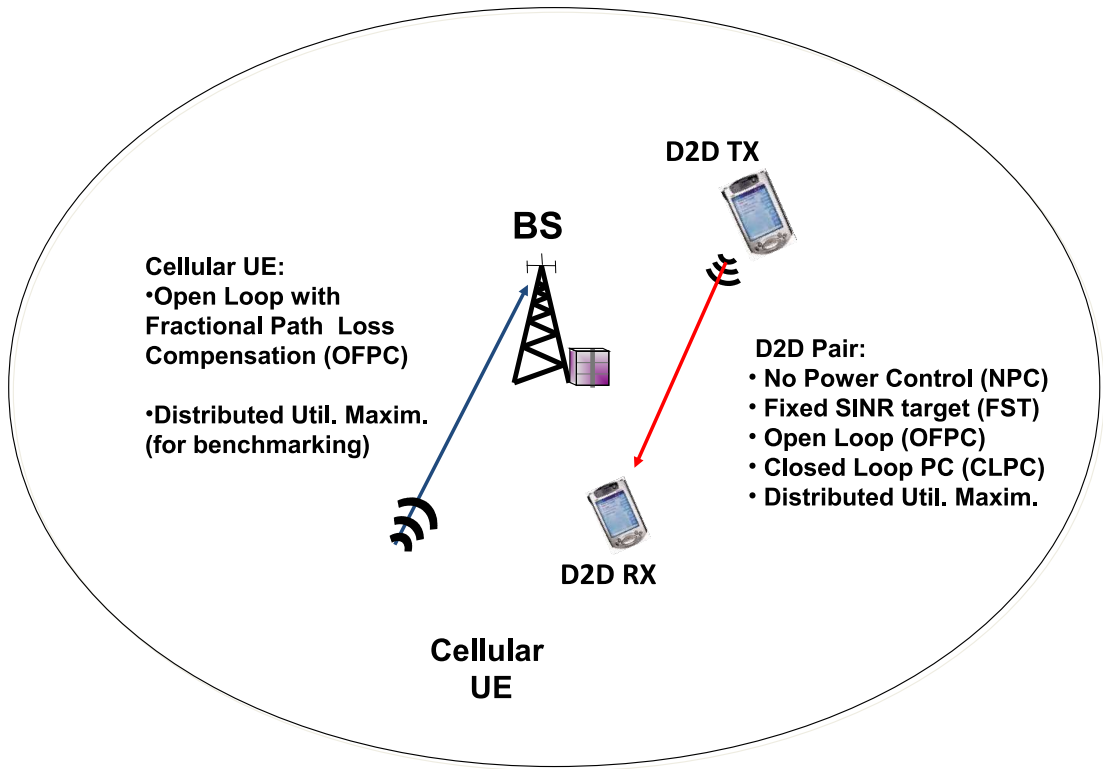}
\caption{The UE communicating in cellular mode with its serving BS
uses the de facto standard LTE OFPC. For the D2D link,
we study various power control strategies, including no power control
(i.e.\ fixed transmit power), fixed SNR target, open loop with fractional
path loss compensation and closed loop that can all
be easily deployed using the flexible standard LTE PC tool box.
In our hybrid scheme, the cellular UEs use LTE OFPC, while the D2D pairs
use utility maximizing PC.
For benchmarking purposes, both the cellular and the D2D users use utility maximization.
}
\label{Fig:PowC}
\end{center}
\end{figure}

For UEs communicating in cellular mode with their respective serving base
stations, OFPC provides a well proven alternative, typically used in practice. It avoids the
complexity and overhead associated with the dynamic offset of the CL scheme, but makes
use of the fractional path loss compensation balancing between overall spectrum efficiency
and cell edge performance \cite{Arne}.
Figure~\ref{Fig:PowC} illustrates the PC options for the D2D link, while we assume
that the cellular link employs the de facto standard LTE fractional path loss compensating
power control scheme.

The PC options available in the integrated cellular and D2D environment are
summarized by Figure \ref{Fig:PowC}.
For cellular users, the LTE OFPC scheme is a viable option, while for D2D users
we are interested in the
performance of various PC alternatives, including those based on the LTE 'tool box' and
utility maximization.
We use the term 'hybrid power control' for the case when cellular UEs
use LTE OFPC, while D2D users use the distributed PC scheme.
As we will see, for benchmarking purposes,
we will allow all (cellular and D2D) users to use the utility
maximizing scheme.

\section{System Model}
\label{Sec:Model}
In order to derive a reference (benchmarking) scheme for network assisted D2D communications,
we model the hybrid cellular-D2D network as a
set of $L$ transmitter-receiver pairs.
A transmitter-receiver pair can be a cellular UE transmitting data to its
serving BS or a D2D pair communicating in cellular uplink spectrum.
D2D {\em candidates} are source-destination pairs in the
proximity of each other that may communicate in direct mode, depending on the
MS decision that is part of the RRM algorithm that is investigated in Section~\ref{Sec:MS}.

The network topology is represented by a directed graph with links labelled with $l=1, \dots , L$ indexing the transmitter-receiver pairs in the network.
Any transmitter, i.e. either cellular or D2D transmitter, operating in the link $l$ is assumed to have always data to send
to the corresponding receiver at a transmission rate $s_l$.
Associated with each link $l$ is a function $u_l(\cdot)$, which describes the {\em utility} of
communicating at rate $s_l$.
The utility function $u_l$ is assumed to be increasing
and \textit{strictly concave}, with $u_l \rightarrow -\infty$ as $s_l \rightarrow 0^+$.
We let $ \mathbf{c}=[c_l]$ denote the vector of link capacities,
which depend on the system bandwidth $W$, the achieved SINR of the links ($\gamma_l$)
as well as the
specific modulation and coding schemes used for the communication.
A feasible rate vector $\mathbf{s}$ must fulfill the following set of constraints:
\[
\mathbf{s}\preceq\mathbf{c}(\mathbf{p}) ,\qquad \mathbf{s}\succeq 0.
\]
In this formulation, it is convenient to look at the $\mathbf{s}$ vector as the
vector of the rate {\em targets} directly derived from a corresponding vector of SINR {\em targets}, while the capacity vector $\mathbf{c}$ depends on the specific powers $\mathbf{p}$ selected by the transmitters. Specifically, each link can be seen as a Gaussian channel with Shannon-like capacity
\begin{equation}
c_l(\mathbf{p})= W \log_2 \big(1+ K\gamma_l(\mathbf{p})\big)
\label{capacity}
\end{equation}
which represents the maximum rate that can be achieved on link $l$, where $K$ models the SINR-gap reflecting
a specific modulation and coding scheme and $\gamma_l(\mathbf{p})$ represents the SINR perceived at the receiver of link $l$. With no loss of generality, we assume $K=1$ in the following.

Let $G_{lm}$ denotes the effective link gain between the transmitter of pair $m$
and the receiver of pair $l$ (including the effects of path-loss and shadowing
\footnote{We assume that the G matrix is obtained after Layer-1 filtering, that is typically used for
open loop power control, mobility management and other purposes in LTE \cite{Arne}.})
and let $\sigma_l$ be the thermal noise power at the receiver of link $l$,
and $P_l$ be the transmission power.
The SINR of link $l$ is
\begin{equation}
\gamma_l(\mathbf{p})= \frac{G_{ll}P_l}{\sigma_l + \displaystyle \sum_{m\neq l}{G_{lm}P_m}}
\label{gamma}
\end{equation}
where $ \mathbf{p}=[P_1, ..., P_L]$ is the power allocation vector, and $\sum_{m\neq l}{G_{lm}P_m}$
is the interference experienced at the receiver of link $l$.
Equation \eqref{gamma} can also be written as
\begin{equation}
\gamma_l(P_{l_{Rx}}^{tot}, P_l, G_{ll})= \frac{G_{ll}P_l}{(P_{l_{Rx}}^{tot} -  G_{ll}P_l)  }
\label{gamma2}
\end{equation}
where $P_{l_{Rx}}^{tot}$ represents the total received power
(including $\sigma_l$)
measured by the receiver of link $l$.
Hence, the SINR in \eqref{gamma2} can be computed by the receiver-$l$ without direct knowledge of any of the channel gains,
except the one related to its corresponding transmitter-$l$.
For the ease of notation, in the following we adopt
$\gamma_l(\mathbf{p})$ to indicate the SINR measured at receiver-$l$.

\section{The SINR Target Setting and Power Control Problem}
\label{Sec:SINR}
\subsection{Formulating the D2D Power Control Problem}
In this section we assume that MS has already been performed for the D2D candidates and a RA algorithm
has already assigned to cellular and D2D links certain RBs for communication.
From the concept of D2D communications reusing cellular spectrum, a given RB may be used
by multiple cellular and D2D transmitters even within the same cell, thus causing intracell interference.
In this section we focus on handling such interference by properly setting the SINR
targets and allocating transmit powers, while the MS and RA problems
that determine the specific cellular and D2D transmitters that share a given RB
are considered to be already been performed. MS and RA are investigated in more details in Sections \ref{Sec:MS} and \ref{Sec:Heuristic}.

For the set of interfering links sharing the same RB and thereby
causing interference to one another,
we formulate the problem of target rate setting and PC as:
\begin{equation}
\begin{array}{ll}
\displaystyle \Max_{\mathbf{p},\mathbf{s}} & \sum_l u_l(s_l) - \omega \sum_l P_l \\
\text{subject to} & s_l\leq c_l(\mathbf{p}) ,\quad \forall l, \\
& \mathbf{p},\mathbf{s}\succeq 0
\end{array}
\label{problem1}
\end{equation}
which aims at maximizing the utility
while taking into account
the transmit powers by means of a predefined weight $\omega \in (0,+\infty) $
\cite{Papa:06},
so as to both increase
spectrum efficiency and reduce the sum power consumption
over all transmitters sharing a specific RB. Constraints of Problem \eqref{problem1} formally ensure that the rate allocation does not exceed the link capacities that in turn depends on the transmit powers on the given RB.
As shown in Section \ref{Sec:Decompose}, Problem \eqref{problem1} can be decomposed into two separate problems (Problem-I and Problem-II) that need to be executed recursively until convergence to the optimum of Problem \eqref{problem1}. Specifically, Problem-I selects the transmit rate target, while Problem-II selects the transmit power that fulfills the desired transmit rate target, i.e. the SINR target. As such, Problem-I and Problem-II resemble an outer-loop and an inner-loop mechanism respectively, where the inner-loop power allocation ensures
that the target rate $\mathbf{s}$ reduces to the optimal capacity vectors ($\mathbf{c}$) at convergence of the outer-inner loop routine.
\vspace{-2mm}
\subsection{Convexifying the Problem of Equation (\ref{problem1})}
\label{Sec:Convex}
Before presenting the decomposition approach, it is important to note that Problem \eqref{problem1} is not convex in its original formulation. However, by 
appealing to the results presented in \cite{Papa:06} and \cite{Belleschi:09},
Problem \eqref{problem1} can be converted into the following equivalent form:
\begin{equation}
\begin{array}{ll}
\displaystyle \Max_{\mathbf{\tilde{s}},\mathbf{\tilde p}}  & \sum_l u_l(e^{\tilde{s}_l}) - \omega \sum_l e^{\tilde{P}_l} \\
\text{subject to} & \log(e^{\tilde{s}_l})\leq \log\left(c_l(e^\mathbf{\tilde p})\right) \quad \forall l,
\end{array}
\label{problem2}
\end{equation}
where $s_l\leftarrow e^{\tilde{s}_l}$ and $P_l \leftarrow e^{\tilde{P}_l}$. 
The transformed Problem~\eqref{problem2} is proved to be convex (now in the $\tilde{s}_l$-s and $\tilde{P}_l$-s)
since 
the utility functions $u_l(\cdot)$
are selected to be
$(log,x)$-concave over their domains \cite{Papa:06}.
In this paper we use $u_l(x) \triangleq \text{ln} (x), \forall l$.
Under this condition, we can solve Problem~\eqref{problem2} to optimality by means of an iterative algorithm
where the $\tilde{s}_l$-s (or equivalently the SINR targets) are set by an outer-loop.
The transmit powers $\tilde{P}_l$-s that meet the particular SINR targets
(set in each outer-loop cycle)
are in turn set by a Zander type iterative SINR target following inner-loop \cite{Zander:92}.
This separation of the setting of the SINR targets and corresponding power levels
are detailed in the next Section.
\section{A Decomposition Approach to the SINR Target Setting and Power Control Problem}
\label{Sec:Decompose}
\subsection{Formulating the Decomposed Problem}
We now reformulate Problem~\eqref{problem2} as a problem
in the user rates $\mathbf{\tilde{s}}$ (\underline{Problem-I}),
which, due to the convexification, can be solved for a given power allocation ($\mathbf{\tilde{p}}$).
Note that the target rate vector $\mathbf{\tilde{s}}$ can be uniquely mapped
to a target SINR vector $\boldsymbol{\gamma}^{tgt}$ as it will be shown later.
We define \underline{\textbf{Problem-I}} as:
\begin{equation}
\begin{array}{ll}
\displaystyle \Max_{\mathbf{\tilde s}}  & \nu(\mathbf {\tilde s} )\\
\text{subject to} &  \mathbf {\tilde s} \in \mathbf{ \tilde S}
\end{array}
\label{problem3}
\end{equation}
where $ \mathbf{\tilde S}= \{\mathbf{\tilde s}|\log(e^{\tilde{s_l}})\leq \log(c_l(e^{\mathbf{\tilde p}})),\forall l \}$
represents the set of feasible rate vectors that, for a given power vector $\mathbf{\tilde p}$,
fulfill the constraints of Problem~\eqref{problem2}.

Comparing \eqref{problem2} and \eqref{problem3}, it follows that
the objective function in \eqref{problem3} is defined as
$\nu(\mathbf{\tilde s}) \triangleq \sum_l u_l(e^{\tilde{s_l}}) - \varphi(\mathbf{\tilde p})$,
where $\varphi(\mathbf{\tilde p}) \triangleq \omega \sum_l e^{\tilde P_l}$
represents the cost in terms of the total transmit power
for realizing a given target rate $\mathbf{\tilde s}$.
Accordingly, we denote with $\varphi^\star(\mathbf{\tilde p}) \triangleq \omega \sum_l e^{\tilde P^\star_l}$
the cost of achieving the optimum rates $\mathbf{\tilde{s}^\star}$ that solve the utility maximization
Problem \eqref{problem3}. 

Therefore, \underline{\textbf{Problem-II}}, for a given $\mathbf{\tilde{s}}$ vector, can be formulated as
\begin{equation}
\begin{array}{ll}
\displaystyle \Min_{\mathbf{\tilde p}}  & \omega \sum_l e^{\tilde P_l}\\
\text{subject to} &  \log( e^{\tilde s_l} ) \leq \log\left(c_l(e^{\mathbf{\tilde p}})\right)\quad \forall l.
\end{array} 
\label{problem4}
\end{equation}
%
Solution approaches to Problem-I and Problem-II are proposed in the next subsection.
\subsection{Solving the Rate (SINR Target) Allocation Problem}
Provided that the objective function $\nu(\mathbf{\tilde s})$  in \eqref{problem3} is concave
and differentiable 
we can determine the optimal $\mathbf {\tilde s^\star}$ by means of
\emph{projected gradient} iterations, with a fixed predefined step~$\epsilon$:
\begin{equation}
{\tilde s_i}^{(k+1)} = \max\left[0,\text{}{\tilde s_i}^{(k)} + \epsilon \nabla_i \nu(\mathbf {\tilde s}^{(k)} )\right]   \quad \forall i,
\label{rate}
\end{equation}
where
\begin{equation}
\nabla_i \nu(\mathbf {\tilde s})= \frac{\partial}{\partial \tilde s_i} \Big[ \sum_l u_l(e^{\tilde{s_l}}) - \varphi^\star(\mathbf{\tilde p}) \Big]
                                = {u_i}'(e^{\tilde s_i}) e^{\tilde s_i}  -  \frac{\partial}{\partial \tilde s_i} \Big[ \varphi^\star (\mathbf{\tilde p}) \Big].
\label{GRADIENTE}
\end{equation}
To compute \eqref{GRADIENTE},
we first need to find $\varphi^\star (\mathbf{\tilde p})$
by solving the primal Problem-II~\eqref{problem4}.
Since it is convex in $\mathbf{\tilde p}$, it can be conveniently
solved by \emph{Lagrangian Decomposition} as follows.
Let $\boldsymbol{\lambda}$ be the Lagrange multipliers (dual variables) for the constraints in \eqref{problem4}
and form the Lagrangian function:
\begin{equation}
\mathcal{L}(\boldsymbol{\lambda},\mathbf{\tilde p}) = \omega \sum_l e^{\tilde P_l} + \sum_l \lambda_l \big[\log\big( e^{\tilde s_l}\big) - \log\left(c_l(e^{\mathbf{\tilde p}})\right) \big].
\label{Lagrangian}
\end{equation}
The Lagrangian dual problem of Problem-II is given by:
\begin{equation}
\begin{array}{ll}
\displaystyle \Max_{\boldsymbol{\lambda}}  & [\mathcal{L}(\boldsymbol{\lambda}) = \displaystyle \min_{\mathbf{\tilde p}} \, \mathcal{L}(\boldsymbol{\lambda},\mathbf{\tilde p})]\\
\text{subject to} &  \boldsymbol{\lambda}\succeq 0.\\
\end{array}
\label{DualPB}
\end{equation}
Since the original problem is convex, if regularity conditions hold 
the solution of Problem~\eqref{DualPB} corresponds to the solution of Problem~\eqref{problem4},
i.e.\ $\mathcal{L}(\boldsymbol{\lambda^\star})= \varphi^\star (\mathbf{\tilde p})$.
Assuming that ($\boldsymbol{\lambda^\star}, \mathbf{\tilde {p}^\star}$)
represents the optimum solution of Problem-II~\eqref{problem4},
we are now in the position to calculate
$\varphi^\star (\mathbf{\tilde p})$ from \eqref{Lagrangian}:
\[
\varphi^\star (\mathbf{\tilde p}) = \sum_l \big[ \omega e^{\tilde P_l^\star} - \lambda_l^\star \log\left(c_l(e^{\mathbf{\tilde p^\star}})\right) \big] + \sum_l \lambda_l^\star \log( e^{\tilde s_l} ) \quad \text{and} \quad \frac{\partial}{\partial \tilde s_i}[\varphi^\star (\mathbf{\tilde p}) ]= \lambda_i^\star.
\]
Recalling \eqref{GRADIENTE}, we have:
%
\begin{equation}
\nabla_i \nu(\mathbf {\tilde s}) = {u_i}'(e^{\tilde s_i}) e^{\tilde s_i}  - \lambda_i^\star
                                  = e^{\tilde s_i}[{u_i}'(e^{\tilde s_i}) - \frac{\lambda_i^\star}{e^{\tilde s_i}}]
                                ={s_i}[{u_i}'({s_i}) - \frac{\lambda_i^\star}{{s_i}}],
\end{equation}
The final target rate update is:
\begin{eqnarray}
{s_i}^{(k+1)} =
e^{{\tilde s_i}^{(k+1)}} = {s_i}^{(k)} \exp\left(\epsilon \nabla_i \nu(\mathbf {\tilde s}^{(k)} )\right). \nonumber
\end{eqnarray}
Combining the above with \eqref{GRADIENTE}, we can write the SINR target
setting rule in the following form:
\begin{equation}\boxed{
{s_i}^{(k+1)} = {s_i}^{(k)} \exp\Bigg(\epsilon \, {s_i}^{(k)} \, \Big[{u_i}'({s_i}^{(k)}) - \frac{\lambda_i^\star ({s_i}^{(k)}) }{({s_i}^{(k)})}\Big]\Bigg)}
\label{RateUpdate}
\end{equation}
Equation \eqref{RateUpdate} dictates the outer-loop mechanism for a certain transmitter $i$. Specifically, at any iteration $(k+1)$, Equation \eqref{RateUpdate} determines the rate (and hence the SINR) that should be targeted during the next inner-loop PC\footnote{We draw a box around equations that need to be implemented
by a receiver or transmitter node, as will be summarized in Figure 3.}.
Following the decomposition approach,
Equation \eqref{RateUpdate} requires the knowledge of the Lagrange multipliers $\lambda_i^\star$
associated with Problem-II, which can be found by solving the PC problem
associated with the $(k)$-th outer-loop iteration.
We consider this specific problem in the next section.
\subsection{Solving the Power Allocation Problem for a given SINR Target}
The inner-loop PC problem (Problem-II) takes as an input a certain SINR target that can be easily derived from Equation \eqref{RateUpdate}.
Given $\mathbf{\tilde s}^{(k)} \in  \mathbf{\tilde S}$ ,
the constraints in \eqref{problem4} correspond to require that the SINR-s of the links
exceed a target value, i.e.\
\[
\log\left( e^{\tilde s_l}\right) \leq \log\left(c_l(e^{\mathbf{\tilde p}})\right) \quad \Leftrightarrow  \quad {\gamma_l}(\mathbf{p}) \geq {\gamma_l}^{tgt}(\mathbf{\tilde s}^{(k)}) \quad \forall l,
\]
where  $\gamma_l(\mathbf{p})$ is defined in \eqref{gamma}, and
\begin{equation}
\boxed{
\gamma_l^{tgt}\left({\tilde s_l}^{(k)}\right) \triangleq 2^{ \frac{e^{\tilde s_l}}{W}} -1.}
\label{gammaTGT}
\end{equation}
Therefore, Problem~\eqref{problem4} can be rewritten as:
\begin{equation}
\begin{array}{ll}
\displaystyle \Min_{\mathbf{\tilde p}}  & \omega \sum_l e^{\tilde P_l}\\
\text{subject to} &  \gamma_l(e^\mathbf{\tilde p}) \geq \gamma_l^{tgt}\left(\tilde s_l\right) \quad \forall l, \\
\end{array}
\label{problem5}
\end{equation}
and solved with an iterative closed-loop PC scheme \cite{Zander:92}:
\begin{equation} \boxed{
{P_l}^{(t+1)}= \frac{\gamma_l^{tgt}(\tilde s_l)}{\gamma_l\big(\mathbf{p}^{(t)}\big)} {P_l}^{(t)}.
\label{power}}
\end{equation}
Thus, for a given $\gamma_l^{tgt}(\tilde s_l)$,
the PC inner-loop \eqref{power} sets the transmit
powers for each transmitter at step $(t+1)$,
provided that the transmitter is aware of the SINR $\gamma_l\big(\mathbf{p}^{(t)}\big)$ measured at the receiver
in the previous step.
\vspace{-3mm}
\subsection{Determining the $\lambda_i^\star$-s}
We can now determine the $\lambda_i^\star$-s for the outer-loop update \eqref{RateUpdate}
by exploiting the relationship between the optimal $\mathbf{p}^\star$ and the associated
Lagrange multipliers $\lambda_i^\star$-s.
To this end, we rewrite the constraints in \eqref{problem5} as:
\begin{eqnarray}
\frac{G_{ll}P_l}{\sigma_l + \displaystyle \sum_{m\neq l}{G_{lm}P_m}} - \gamma_l^{tgt} \geq 0 \quad \Rightarrow 
P_l - \gamma_l^{tgt} \sum_{m\neq l} \frac{G_{lm}}{G_{ll}}P_m - \frac{\gamma_l^{tgt} \sigma_l}{G_{ll}} \geq 0 \quad \forall l.
\end{eqnarray}
Furthermore, let $\mathbf{H} \in \mathbb{R}^{LxL}$ and $\mathbf{\eta} \in \mathbb{R}^{L}$ be defined as follows:
\begin{eqnarray}
\begin{array}{lr}
\mathbf{H}=[h_{lm}] \triangleq  \left\{
\begin{array}{l l}
\label{eq:H}
     -1                                      & \text{if $l=m$}\\
     \gamma_l^{tgt} \frac{G_{lm}}{G_{ll}}   & \text{if $l\neq m$}\\
  \end{array} \right.  \\
  \\
\boldsymbol{\eta}=[\eta_l] \triangleq \Big[\frac{\gamma_l^{tgt} \sigma_l}{G_{ll}} \Big].
\end{array}
\end{eqnarray}
Using this notation, we can reformulate Problem \eqref{problem5} as the following \underline{\emph{Linear Programming}} (LP) problem:
\begin{equation}
\begin{array}{ll}
\displaystyle \Min_{\mathbf{p}}  & \omega \mathbf{1}^T \mathbf{p}\\
\text{subject to} &  \mathbf{Hp} \preceq - \boldsymbol{\eta}; \quad \mathbf{p} \succeq 0, \\
\end{array}
\label{problem6}
\end{equation}
with the corresponding \underline{\textit{Dual Problem}}
\begin{equation}
\begin{array}{ll}
\displaystyle \Max_{\mathbf{\lambda^{\text{\tiny(LP)}}}}          & \boldsymbol{\eta}^T \boldsymbol{\lambda}^{\text{\tiny(LP)}}\\
\text{subject to} &  \mathbf{H}^T \boldsymbol{\lambda}^{\text{\tiny(LP)}} \succeq - \omega \mathbf{1}; \quad \boldsymbol{\lambda}^{\text{\tiny(LP)}}\succeq 0 \\
\end{array}
\label{problem7}
\end{equation}
which is necessary to compute the Lagrange multipliers in Equation~\eqref{RateUpdate} for the rate update.\\
As it is shown in Appendix A, the inequality constraints in \eqref{problem7} can be rewritten explicitly as:
\begin{equation}
\frac{ {\lambda}_l^{\text{\tiny(LP)}}}{\omega} - \sum_{k \neq l} \frac{G_{kl}}{G_{kk}} {\gamma}_k^{tgt} \frac{ {\lambda}_k^{\text{\tiny(LP)}}}{\omega} \leq 1 ,\quad \forall l.
\label{inequality}
\end{equation}
As it is shown in Appendix B, by defining
\begin{equation}
\mu_l \triangleq \frac{ \lambda_l^{\text{\tiny(LP)}}}{\omega} \frac{\gamma_l^{tgt} \sigma_l}{G_{ll}} = \frac{ \lambda_l^{\text{\tiny(LP)}}}{\omega}  \eta_l,
\label{mu}
\end{equation}
Equation~\eqref{inequality} can be interpreted as an SINR requirement, i.e.\
\begin{equation} \boxed{
\gamma_l^{cc} (\boldsymbol{\mu}) \triangleq \frac{\mu_l G_{ll}}{\sigma_l + \displaystyle \sum_{k\neq l} G_{kl} \frac{\sigma_l}{\sigma_k}\mu_k} \leq \gamma_l^{tgt}, \quad \forall l.}
\label{SIR}
\end{equation}

Therefore, Problem \eqref{problem7} can be reformulated as:
\begin{equation}
\begin{array}{ll}
\displaystyle \Max_{\boldsymbol{\mu}}  & \omega \mathbf{1}^T \boldsymbol{\mu}\\
\text{subject to} & {\gamma_l}^{cc} \leq {\gamma_l}^{tgt}, \quad \forall l; \quad \boldsymbol{\mu}\succeq 0 \\
\end{array}
\label{problem8}
\end{equation}
where the solution  $\boldsymbol{\mu}$ can be computed according to the following distributed closed-loop PC similarly to Equation~\eqref{power}
\begin{equation}\boxed{
\mu_l^{(t+1)} = \frac{\gamma_l^{tgt}}{\gamma_l^{cc}(\boldsymbol{\mu}^{(t)})} \mu_l^{(t)} \quad \forall l .}
\label{muUpdate}
\end{equation}
%
Equation~\eqref{muUpdate} can be interpreted as a reverse link PC problem that is executed in the control channel between the receiver and the transmitter of link $l$. Specifically, the receiver-$l$ adapts its transmitting power $\mu_l$ according to Equation~\eqref{muUpdate}, while the transmitter-$l$ measures the experienced SINR $\gamma_l^{cc}$ in the corresponding control channel.\\
Once the iterative procedure \eqref{muUpdate} converges to the optimum  $\boldsymbol{\mu}^\star$,
the optimal dual variables $\boldsymbol\lambda^{\star\text{\tiny(LP)}}$ can be retrieved from Equation~\eqref{mu} as
\begin{equation}
\lambda_l^{\star\text{\tiny(LP)}} = \omega \mu_l^\star \eta_l^{-1}, \quad \forall l.
\label{lambdaLP}
\end{equation}
The original nonlinear PC  problem \eqref{problem4} and the corresponding LP formulation \eqref{problem6} are equivalent
in the sense that there exists the following specific relation between their optimal solutions
$(\mathbf{\tilde p}^{\star}, \boldsymbol{\lambda}^{\star})$
and $(\mathbf{p}^{\star}, \boldsymbol{\lambda}^{\star\text{\tiny(LP)}})$:
\begin{equation}
\begin{array}{l}
P_l^{\star} = e^{{\tilde P_l}^{\star}}  \qquad \forall l\\
\lambda_l^{\star} = \log(1 + \gamma_l^{tgt}) \frac{1 + \gamma_l^{tgt}} {\gamma_l ^{tgt}} P_l^{\star} \log(2) \lambda^{\star\text{\tiny(LP)}} \qquad \forall l.
\end{array}
\label{relation}
\end{equation}
Hence, once both $P_l^{\star}$ and $\mu_l^{\star}$ are achieved
by means of Equations~\eqref{lambdaLP} and \eqref{relation}, we are able to compute $\boldsymbol{\lambda}^\star$  as
\begin{equation}\boxed{
\lambda_l^{\star} = \log(1 + \gamma_l^{tgt}) \frac{1+ \gamma_l^{tgt}}{\gamma_l^{tgt}} {P_l}^{\star} \log(2) \omega \mu_l^\star \frac{G_{ll}}{\sigma_l \gamma_l^{tgt}} \qquad \forall l,}
\label{LambdaOPT}
\end{equation}
Equation \eqref{LambdaOPT} is then used to update the user rates in Equation~\eqref{RateUpdate}.

\subsection{Summary}
\begin{figure}[t!]
\begin{center}
\includegraphics[width=0.85\hsize]{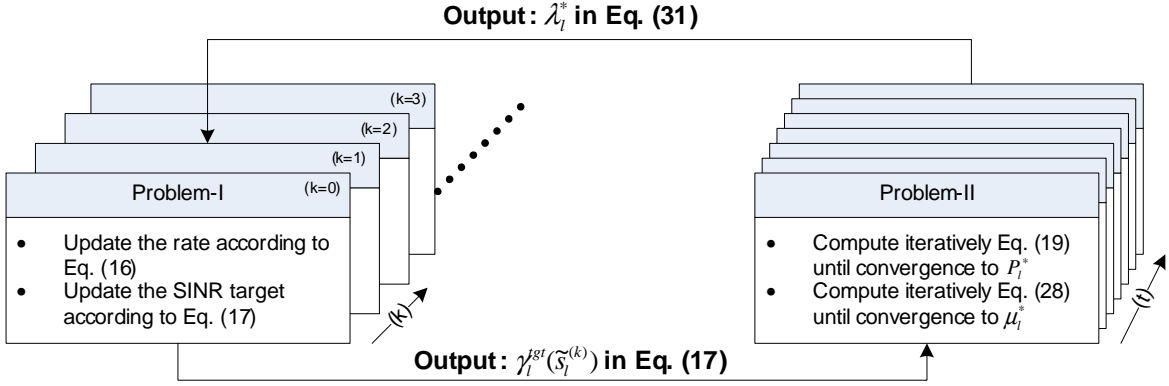}
\caption{Machinery of the distributed utility maximization algorithm presented in Section \ref{Sec:Decompose}.
The algorithm can be executed by any transmitter-receiver pair in the network, i.e.\ both D2D and cellular.
At convergence, the outer-loop provides the optimal
SINR targets, i.e.\ transmit rates, while the inner-loop provides the optimal associated transmit power levels for any transmitter.
In a real-world scenario, Equation \eqref{RateUpdate} is computed by any transmitter and serves as an input (Equation \eqref{gammaTGT}) to the inner-loop PC, i.e. Equation \eqref{power} and \eqref{muUpdate}.
In turn, the inner-loop PC dictates the rate update at the next iteration through Equation \eqref{LambdaOPT}.
}
\label{Fig:Loops}
\end{center}
\end{figure}

In this section, we have explored an outer-inner loop iterative solution
for the convex optimization Problem \eqref{problem2}.
The basic idea is to decompose Problem \eqref{problem2} into separate
subproblems in $\mathbf{\tilde{s}}$ (Problem-I \eqref{problem3})
and $\mathbf{\tilde p}$ (Problem-II \eqref{problem4}).
For each link $l$, Problem-I and Problem-II operate in concert as show in Figure \ref{Fig:Loops}.
Problem-I is in charge of the outer-loop iterations, while Problem-II deals with the inner-loop PC.
More specifically, the solution of Problem-I at step $(k)$, i.e. $\gamma_l^{tgt}\left({\tilde s_l}^{(k)}\right)$,
serves as input of Problem-II that is executed until convergence to $\mu_l^\star$ and $P_l^\star$.
In turn, Problem-II outputs $\lambda_l^{\star}$ that is used by a new instance of Problem-I at step $(k+1)$.
It is important to note that given the constraints of Problem-I, Problem-II is always provided
with a set a feasible SINR target that can be achieved in a finite number of iterations by finite values of $\mathbf{p}$.
In other words, the solutions of Problem-II in Equation \eqref{RateUpdate} always
move within the rate feasibility region $\mathbf{\tilde{S}}$,
provided that the step size $\epsilon$ is small enough.
In a setting at step $(k=0)$, the outer-loop can be initiated with a low feasible SINR target vector
that allows the inner-loop to determine in a finite number of iterations
the finite transmit power levels $\mathbf{p^\star}$ and the corresponding
$\mathbf{\lambda}^\star$.

\section{The Mode Selection (MS) and Resource Allocation (RA) Problem}
\label{Sec:MS}
\subsection{Basic Considerations for Mode Selection and Resource Allocation}
While cellular UEs communicate with their respective serving BS,
D2D-capable UEs may communicate either in direct mode with their respective D2D pairs or in cellular mode with the serving BS.
In the direct mode case, D2D transmitters are allowed either to reuse cellular RBs, i.e. D2D reuse mode, or
allocated orthogonal (dedicated) RBs, i.e. D2D dedicated mode. In the latter case, the reuse gain of D2D communications is not harvested.

On the other hand, when a D2D-capable UE communicates in cellular mode,
D2D communication reduces to the ordinary cellular communication and
RA follows the legacy OFDMA allocation strategy, i.e.\ RBs are allocated orthogonally between all UEs.
Therefore, three different communication modes can be considered for D2D communications:
D2D mode with dedicated resources, D2D mode reusing cellular resources and cellular mode.
We note that when the D2D candidate pairs communicate in cellular mode,
downlink resources need to be allocated for the BS-D2D receiver link. For the sake of ease, downlink resource
usage is not modeled in this paper.

We now consider a cellular system with $N$ cellular UEs and
$M$ D2D transmitters and corresponding $M$ D2D receivers belonging to the sets $\mathcal{N}$
and $\mathcal{M}$ respectively such that the total number of users in a cell is $L=N+M$.
We denote with $x_{l,j}(q)$ that indicate whether a transmitter-receiver pair $l$ is assigned to RB-$j$
in communication mode $q$, where $(q=0)$ denotes cellular mode and $(q=1)$ the
D2D direct mode.
By definition, any cellular UE $n\in \mathcal{N}$ always transmits in the cellular mode $(q=0)$,
while a D2D candidate can be forced either to operate
using the direct link $(q=1)$,
or the cellular mode $(q=0)$, or adaptively switch between the direct and cellular link according to a specific MS algorithm.
With this terminology at hand, we can formulate the resource constraints as follows:
\begin{itemize}
\item \emph{Forced D2D mode}: \\
$x_{m,j}(q) = x_{m,j}(1), \quad \forall m\in \mathcal{M}$ \quad and \quad $\displaystyle \sum_{n\in \mathcal{N}} x_{n,j}(0) \leq 1, \quad \forall j$
\item \emph{Forced cellular mode}:  \\
$x_{m,j}(q) = x_{m,j}(0), \quad \forall m\in \mathcal{M}$ \quad and \quad $\displaystyle \sum_{n\in \mathcal{N}} x_{n,j}(0) +  \sum_{m\in \mathcal{M}} x_{m,j}(0) \leq 1, \quad \forall j$
\item \emph{Adaptive MS}:\\ 
$\sum_q x_{m,j}(q) \leq 1, \quad\forall m\in \mathcal{M}$ \quad and \quad $\displaystyle \sum_{n\in \mathcal{N}} x_{n,j}(0) + \sum_{m\in \mathcal{M}}   x_{m,j}(0) \leq 1, \quad \forall j$
%
\end{itemize}
where the last inequality indicates that a specific D2D candidate pair $m$ can only be
either in D2D or cellular mode when using RB-$j$.
Note that formally a specific D2D candidate pair $m$ is allowed to use some RBs
in D2D mode and other RBs in cellular mode.
\vspace{-2mm}
\subsection{Formulating the Mode Selection and Resource Allocation Problem}
\label{Sec:RA2}
We now formulate the problem of allocating RBs to users
(both cellular UEs and D2D pairs) $l\in L$, and selecting
the appropriate communication mode~($q$) for the D2D pairs
in order to take advantage of the potential proximity.
More specifically,
the RA task is formulated as a single
\textit{cell-based} optimization problem 
that maximizes the overall spectral efficiency for a given power allocation vector.
The spectral efficiency for transmitter-receiver pair $l$ on a given RB-$j$ can be defined as
$\eta_{l,j}=\log_2\left(1 + \frac{G_{l,l,j}P_l}{\sigma + I_{l,j}}\right)$.
Hence, it depends on
the path gain $G_{k,l,j}$ between transmitter-$k$ and receiver-$l$ on the RB-$j$ and
the intracell interference $I_{l,j} = \sum_{k \neq l} P_l \cdot G_{k,l,j}$,
due to the possible RB sharing between D2D pairs and cellular-UEs.

Thus, the user assignment task becomes (\underline{\textbf{Problem-III}}):
\begin{align}
\label{problem9}
\mbox{maximize} & \sum_l \sum_j
\log_2\left(1 + \frac{G_{ll,j}P_l \cdot x_{l,j} (q) }{\sigma + I_{l,j}}\right) & \\
\mbox{subject to} & \sum_l x_{l,j} (0) \leq 1 \quad \forall j  \tag{C1} \label{eqn:C1}\\ \nonumber
& \sum_q x_{l,j}(q) \leq 1,\quad\forall l,j  \tag{C2} \label{eqn:C2} \nonumber\\
& x_{n,j}(1) = 0 \quad \forall n \in \mathcal{N},j \nonumber \tag{C3} \\
& x_{l,j} (q) \in \{0,1\} \tag{C4} \nonumber
\end{align}
The constraints (C1) indicate that each RB can be allocated to at most one user
in cellular mode due to the orthogonality constraint.
Constraints (C2) ensure that to each user is assigned only one of the two possible modes.
By definition, cellular UEs must not be assigned to mode $(q=1)$ (C3).

\section{Heuristic Algorithms to Solve the Resource Allocation Problem}
\label{Sec:Heuristic}
\subsection{The MinInterf Algorithm}
To solve Problem~\eqref{problem9} and to obtain benchmarking results,
we first propose a centralized procedure
based on the full knowledge of the  path loss measurements between all
transmitters and receivers within the cell.
This scheme, that we call MinInterf, exploits the proximity between D2D candidates
for MS, and performs RA that aims at
reducing the intracell interference by minimizing the sum of the harmful path gains
as will be shown in Equation \eqref{sum-gain} and in Algorithm 1.
MinInterf involves two steps.
Firstly, orthogonal resources are allocated to cellular UEs employing legacy
RA schemes.
Next, for each D2D candidate in the cell, MinInterf considers
two possible cases:
\begin{itemize}
\item
\textit{D2D transmission with dedicated resource}.
If there are orthogonal resources left, they can be assigned to the D2D candidate
so that the D2D transmission does not affect others within the same cell.
In this case, the D2D transmitter
selects the best communication mode
(i.e.\ Cellular Mode or D2D Mode)
on the basis of the path gains both towards the D2D receiver
($G_{d2dMode}$) and the BS ($G_{CellularMode}$).
Specifically, if $G_{CellularMode} \leq G_{d2dMode}$,
then the direct mode is preferred. (See Figure \ref{Fig:PPTFig3}.)

\begin{figure}[t!]
\begin{center}
\includegraphics[width=0.80\hsize]{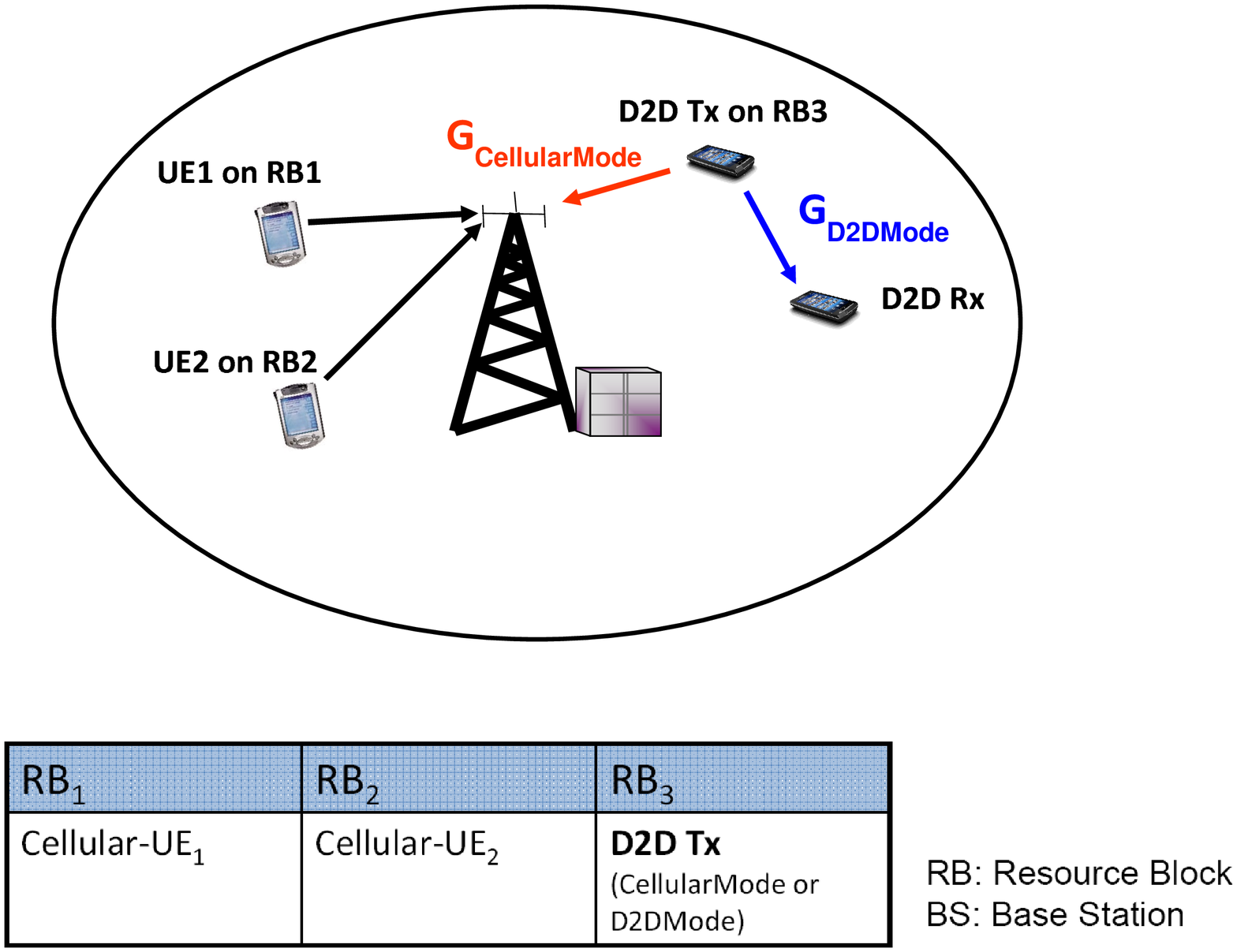}
\caption{An example of a D2D transmission with dedicated resource. The D2D Tx
node selects the transmission-mode (Cellular Mode or D2D Mode)
according to the shadowed path loss measurements towards the D2D Rx
node and towards the BS. If the channel gain between the D2D pair is
higher than the one towards the BS, then the D2D Mode is preferred.}
\label{Fig:PPTFig3}
\end{center}
\end{figure}

\item \textit{D2D transmission with resource reuse (as in Figure \ref{Fig:PPTFig4})}.
When there are no unused RBs in the cell,
the D2D pair must communicate in direct mode (D2D Mode) and reuse RBs.
Sharing resources with other users within the same cell produces intracell interference.
To reduce this intracell interference,
for each resource-$j$ MinInterf considers the sum
\begin{equation}
\mathbf{S}(j)=G_{2Tx\_1Rx , j} + G_{1Tx\_2Rx , j} \quad \text{[dB]}
\label{sum-gain}
\end{equation}
as a measure of the potential interference that assigning the D2D-pair
to resource-$j$ causes. Here $G_{2Tx\_1Rx , j}$ represents the path
gain between the D2D transmitter and the receiver of link(s) already
allocated to resource-$j$, which may be the cellular BS and/or
other D2D receiver(s).
$G_{2Tx\_1Rx , j}$ takes into account the interference that the D2D
pair produces transmitting on RB-$j$.
$G_{1Tx\_2Rx , j}$, on the other hand, is the path gain between the transmitter(s)
already allocated to RB-$j$
(which can be both a cellular-UE and/or other D2D transmitters)
and the receiver of the new D2D pair to be allocated.
$G_{1Tx\_2Rx , j}$ is therefore related to the interference
that the D2D receiver will experience
due to the reuse.
Once expression~\eqref{sum-gain} is computed for each available resource-$j$,
the D2D pair is assigned to that resource corresponding to the minimum value.
(See Figure \ref{Fig:PPTFig4}.)

\begin{figure}[t!]
\begin{center}
\includegraphics[width=0.80\hsize]{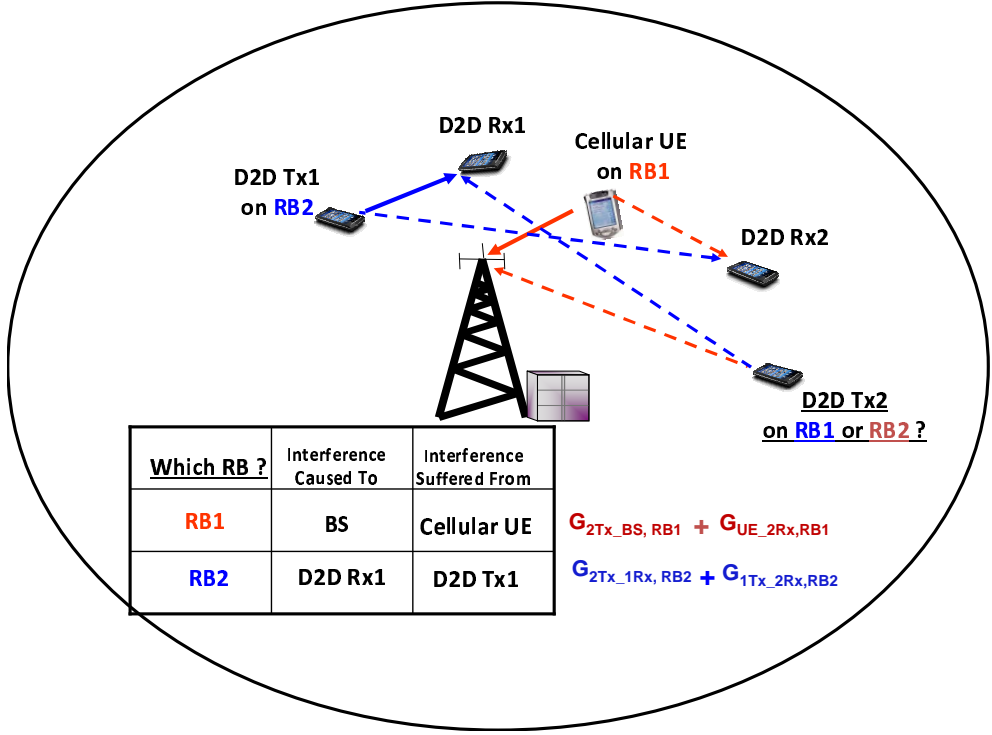}
\caption{An example of a D2D transmission with resource reuse.
D2D Tx node communicates directly to its D2D Rx node sharing
a resource block (RB) with the cellular user UE.
The shared RB is selected in such a way to minimize
an estimate (Equation~\eqref{sum-gain}) of the intracell interference
that D2D communication might perceive (related to the gain $G_{1Tx\_2Rx}$
between the UE and the D2D Rx node) and produce (related to the gain $G_{2Tx\_1Rx}$
between the D2D Tx node and the BS).
}
\label{Fig:PPTFig4}
\end{center}
\end{figure}
\end{itemize}
It is worth noting that the final RA achieved by
MinInterf represents a suboptimal solution of Problem~\eqref{problem9},
nevertheless numerical results show that its interplay with the iterative PC procedure
allows to attain good performance in terms of spectrum and energy efficiency.
\textbf{Algorithm 1} summarizes the main steps of the MinInterf scheme.

\begin{algorithm}[htbp]
\caption{MinInterf}
\begin{algorithmic}
{\small
\STATE Allocate orthogonal resources (RB) to cellular-UEs (using legacy algorithms)
\FOR {Each D2D candidate}
	\IF {there is an orthogonal resource-$l$ left}
	 		\IF {$G_{CellularMode} \leq G_{d2dMode}$}
	 		\STATE	D2D candidate transmits in \textit{D2D-Mode} on resource-$l$
	 		\ELSE
	 		\STATE	D2D candidate transmits in \textit{Cellular-Mode} on resource-$l$
	 		\ENDIF
	\ELSE
        \STATE /* \textbf{Resource Reuse} as in Figure \ref{Fig:PPTFig4} */
	 	\FOR {Each available resource-$j$}
	 	\STATE $\mathbf{S}(j)=[G_{2Tx\_1Rx , j} + G_{1Tx\_2Rx , j}]$
	 	\ENDFOR
	\STATE D2D candidate transmits in \textit{D2D-Mode} on resource-$j$ corresponding to the minimum value of $\mathbf{S}$
	\ENDIF
\ENDFOR
}
\end{algorithmic}
\end{algorithm}

\begin{algorithm}[htbp]
\caption{Balanced Random Allocation (BRA) and Cellular Protection Allocation (CPA)}
\begin{algorithmic}
{\small
\STATE $\rho_j = 0$ for all RB-$j$
\IF {there are cellular UEs in the cell}
    \STATE Allocate orthogonal resources (RB) to cellular-UEs (using legacy algorithms)
    \STATE Set $\rho_j = 1$ for RB:s assigned to UEs
    \STATE /* For \textbf{CPA}: Store $g(j)$, where $g(j)$ is the path gain between cellular
            UE using RB-$j$ and the serving BS */
\ENDIF
\FOR {each D2D candidate}
    \STATE $\rho_{MIN} := \displaystyle \min_{j=1 \dots R} \rho_j$ where $R$ is the total number of resource blocks
    \IF {$\rho_{MIN} == 0$}
            \STATE /* there is an orthogonal resource-$l$ left:
                Schedule D2D on orthogonal resource-$l$ as in Figure \ref{Fig:PPTFig3} */
            \IF {$G_{CellularMode} \leq G_{d2dMode}$}
            \STATE  D2D candidate transmits in \textit{D2D-Mode} on resource-$l$
            \ELSE
            \STATE  D2D candidate transmits in \textit{Cellular-Mode} on resource-$l$
            \STATE /* For \textbf{CPA}: Store $g(l)$, where $g(l)$ is the path gain
                    between the D2D transmitter in cellular mode using RB-$l$ and the serving BS */
            \ENDIF
    \ELSE
        \STATE /* \textbf{Resource Reuse} as in Figure \ref{Fig:PPTFig4} */
        \STATE Select a resource-$j$ out of the resources for which $\rho_j==\rho_{MIN}$ (e.g.\ based on channel state/quality information)
        \STATE /* For the \textbf{CPA} algorithm: Substitute the above by: Pick the resource-$j$ out of the resources for which $\rho_j==\rho_{MIN}$ for which $j=\arg \max g(j)$, where $g(j)$ is the path gain between the cellular transmitter using RB-$j$ and the BS */
        \STATE D2D candidate transmits in \textit{D2D-Mode} on resource-$j$
    \ENDIF
    \STATE Increment $\rho_j$
\ENDFOR
}
\end{algorithmic}
\end{algorithm}

\subsection{Practical MS and RA Algorithms with Limited or No Channel State Information: BRA and CPA}
While MinInterf can serve as a tool to benchmark RA algorithms, it cannot be employed
in practice because it relies on a full $G$ matrix knowledge in the ``Resource Reuse''
branch of the \textbf{Algorithm 1}.
Therefore we seek viable alternatives to MinInterf.
Our first proposed algorithm operates
without any path loss knowledge but keeps
track of the reuse factors $\rho_j$ of each RB as described by the pseudo code of
the ``Balanced Random Allocation'' (BRA, \textbf{Algorithm 2}).
$\rho_j$ is a counter associated with resource-$j$ that counts the number of
intracell transmitters using that resource.
\footnote{We note that BRA can be made completely distributed by skipping the usage of $\rho_j$ in the
algorithm. Simulation results (not shown here) indicate that the impact of skipping $\rho_j$
in BRA is not significant.}

Our second proposed practical algorithm is called ``Cellular Protection Allocation'' (CPA).
CPA takes advantage of the knowledge of the path gains between any cellular transmitter
(i.e. cellular UE or D2D candidate operating in cellular mode)
and the BS, that is available in practice 
due to measurement reports by the UE.
As indicated in the pseudo code of \textbf{Algorithm 2}, a D2D transmitter that
reuses a cellular RB is assigned to the particular RB used by a cellular UE that
has the strongest cellular link. The rationale for this heuristic is that a cellular
UE with a strong cellular connection with its serving BS can be expected to
tolerate intracell interference caused by D2D resource reuse.
\section{Numerical Results}
\label{Sec:Num}

\subsection{Simulation Setup and Parameter Setting}
In this section we consider the uplink (UL) of a 7-cell system,
in which the number of UL physical resource blocks (RB) is 8 (per cell).
We perform Monte Carlo experiments to build statistics over the
performance measure of interests when employing the MinInterf, CPA and
BRA resource allocations together with the utility maximizing,
LTE based or hybrid PC.
In the hybrid scheme, the cellular UEs use the LTE open loop fractional path
loss compensating power control, while the D2D users use the utility maximizing scheme.
In the hybrid scheme, the D2D transmitters are assumed to know their path gains
to the cellular BS (using cellular measurements) and limit their transmit power levels
such that the caused interference at the BS remains under the parameter $I^\star$.

In each cell we drop 6 cellular UEs
that communicate with their respective serving BS using 1 UL RB.
In addition, 6 D2D {\it candidate} pairs are also dropped
in the coverage area of each cell.
For the D2D users
the system may select the D2D mode
or cellular mode to communicate, as described in Section \ref{Sec:MS}.
When a D2D pair uses the cellular mode, the D2D transmitter transmits data to the
BS in the UL band, and the BS sends this data to the D2D receiver in the DL band.
In our study, we do not model the DL transmission, essentially assuming that the
DL resources are in abundance so that we can focus on the UL performance.
When the D2D pair communicates in the direct mode, the D2D transmitter sends
data to the D2D receiver using UL resources.
This case is referred to ``MS'' to emphasize the role of the mode selection
for the D2D candidates.

Since 6 D2D pairs are dropped in addition to the 6 cellular users,
4 of them must use direct mode and overlapping resources with either other D2D direct
mode users or with cellular users.
This is because we assume 8 resources per cell accommodating 12 transmitters
and we assume that cellular users and D2D candidates in cellular mode (that is
transmissions to the cellular BS) must remain orthogonal within a cell.
We refer to this case as the ``MS Reuse'' to highlight that there is a degree of
mode selection freedom for two D2D candidates but cellular resources now must
be {\it reused} by multiple transmitters in each cell.
Intuitively, we expect some SINR degradation on the reused resources, but an
increase in the total rate (and spectrum efficiency) due to more transmissions
per cell.
Distinguishing the two D2D pairs case and the four D2D pairs case allows us to
separate the proximity gain (without reuse gain) from the reuse gain
(expected in the second case).

When the utility maximizing PC (as a reference case) is used,
all (cellular and D2D) transmitters employ the outer-inner loop based power
control. In contrast, when the LTE PC is used, the cellular UEs
use standard LTE open loop fractional path loss compensation (OFPC) method,
whereas we test fixed SNR target, fixed transmit power and the closed loop
based method for the D2D link.

The main simulation parameters are given in Table~\ref{parameters}.
{\small
\begin{table}[t!]
\begin{center}
\caption{Parameters of the 7-cell System Under Study}
\vspace{3mm}
\label{tab:Table1}
\label{parameters}
\begin{tabular}[h]{|l|l|}
\hline
\emph{\textbf{Parameter}}     &       \emph{\textbf{Value}} \\
\hline
System Bandwidth    &   5MHz     \\
\hline
Carrier Frequency   & 2GHz       \\
\hline
Gain at 1 meter distance    & -37dB \\
\hline
Thermal noise $\sigma$          & -114 dBm \\
\hline
Path Loss coefficient & 3.5\\
\hline
Lognormal shadow fading  & 6dB\\
\hline
Cell Radius &   500m\\
\hline
Number of cells & 7\\
\hline
Max Tx Power & 200mW\\
\hline
Min Tx Power & 5e-6W\\
\hline
$P_{IN}$ & -116 dBm \\
\hline
$\alpha$ & 0.8 \\
\hline
Number of RB's requested by users&  1\\
\hline
Max. Number of Outer-Loop iterations &  100\\
\hline
Max. Number of Inner-Loop iterations &  10 \\
\hline
Number of MonteCarlo simulations &  100\\
\hline
Initial power  &    0.01 W\\
\hline
Initial $\gamma^{tgt}$    & 0.2\\
\hline
Initial $\mu$ & 0.01\\
\hline
$\epsilon$ &    0.05\\
\hline
 $\omega$ & 0.01 \dots 10 \\
\hline
Distance between D2D pairs & 50-100m\\
\hline
Maximum interference caused by D2D users at the BS & $I^\star=0.02 \cdot N_0 \dots 500 \cdot N_0$ \\
\hline
\end{tabular}
\end{center}
\end{table}
}

\subsection{Numerical Results}
\begin{figure}[t!]
\begin{center}
\includegraphics[width=0.85\hsize]{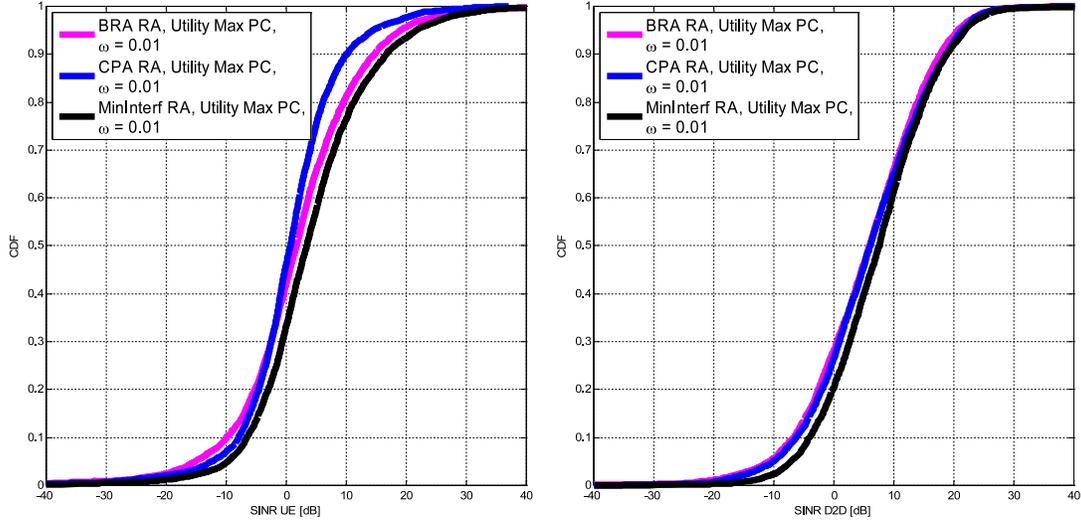}
\caption{
Comparing the performance of the MinInterf, CPA and BRA resource allocations in terms of
SINR distribution of the cellular UEs and D2D pairs when employing the utility maximizing PC
scheme for all (i.e.\ cellular UE and D2D) users.
MinInterf is clearly superior in terms of protecting the cellular UEs in the entire SINR region.
Since CPA tends to reuse resources used by the strong (cell center) UEs,
its SINR performance is somewhat worse in the high SINR region than the BRA.
}
\label{Fig:pairing}
\end{center}
\end{figure}

\begin{figure}[ht!]
\begin{center}
\includegraphics[width=0.85\hsize]{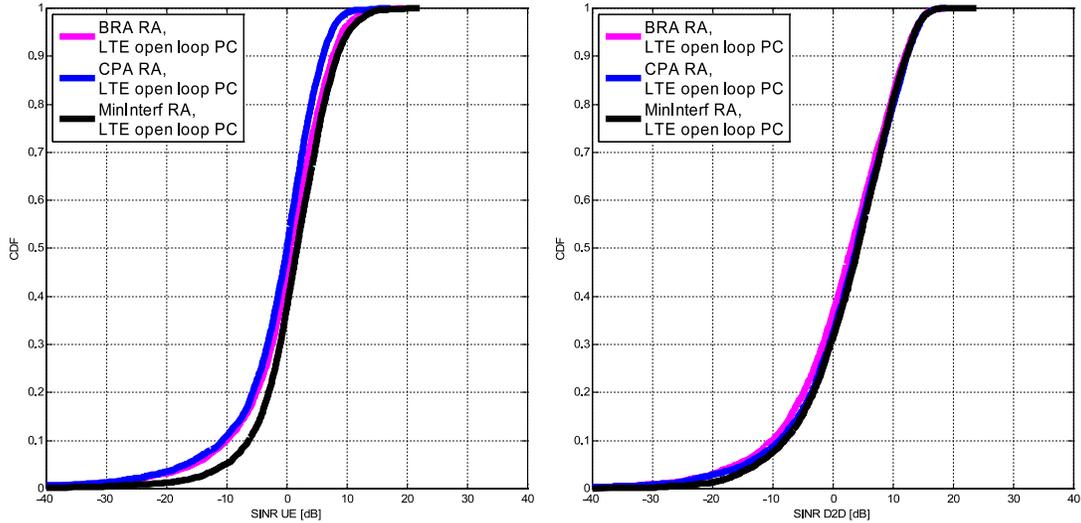}
\caption{
Comparing the performance of the MinInterf, CPA and BRA resource allocations in terms of
SINR distribution of the cellular UEs and D2D pairs when using the LTE open loop PC.
The 3 RA schemes perform quite similarly, but somewhat worse than with
the utility optimal PC, especially in the high SINR region (5-8 dB performance loss
for the 80\%-ile and stronger users).
}
\label{Fig:pairing2}
\end{center}
\end{figure}

Figure \ref{Fig:pairing} and Figure \ref{Fig:pairing2} compare the SINR performance
of MinInterf, CPA and BRA for the cellular UEs and D2D pairs when using
the utility maximizing and the LTE
open loop fractional path loss compensation power control respectively.
Because there are only 8 RBs per cell, at most 2 D2D candidates may choose cellular mode,
while the remaining 4 D2D candidates must use direct mode and reuse a RB for its transmission.
When using the utility maximizing PC (both for the cellular UEs and D2D pairs),
the SINR performance of the MinInterf, CPA and BRA resource allocation schemes is very
similar, except in the high SINR region of the cellular users, that could gain a 6-8 dB
SINR increase with MinInterf as compared to CPA.
Somewhat surprisingly, BRA gets closer
to the performance of MinInterf in this region at the expense of performing a bit worse
for cell edge users than CPA.
The reason for this is that CPA tends to reuse the RBs of
strong (cell center) cellular users.

With the LTE PC, MinInterf shows gains over CPA and BRA for cellular UEs
in the low SINR regime (up to 5 dB gain), essentially
protecting the cell edge UEs from excessive interference from the D2D traffic.
For example, the gain of MinInterf at the
50\% percentile is only around 1-2 dB (see Figure \ref{Fig:pairing2}),
which is somewhat disappointing considering the full path loss matrix requirement of MinInterf.
In scenarios in which the cellular UEs are more far from their respective serving BSs
(not shown here), the gain of MinInterf is greater, but still typically remains under 3 dB difference.
Also, somewhat surprisingly, there is no notable difference between the performance
of the CPA and BRA allocations.
However, comparing the D2D SINR distributions of Figure \ref{Fig:pairing} and Figure \ref{Fig:pairing2},
we observe a significant gain obtained by the utility maximizing scheme in the range of 5-8 dB
throughout the CDF.
These results encourage us to use the simple balanced random -- BRA -- resource allocation scheme in the
remaining of the numerical section and focus on comparing the performance of different PC approaches.
%

\begin{figure}[t!]
\begin{center}
\includegraphics[width=0.85\hsize]{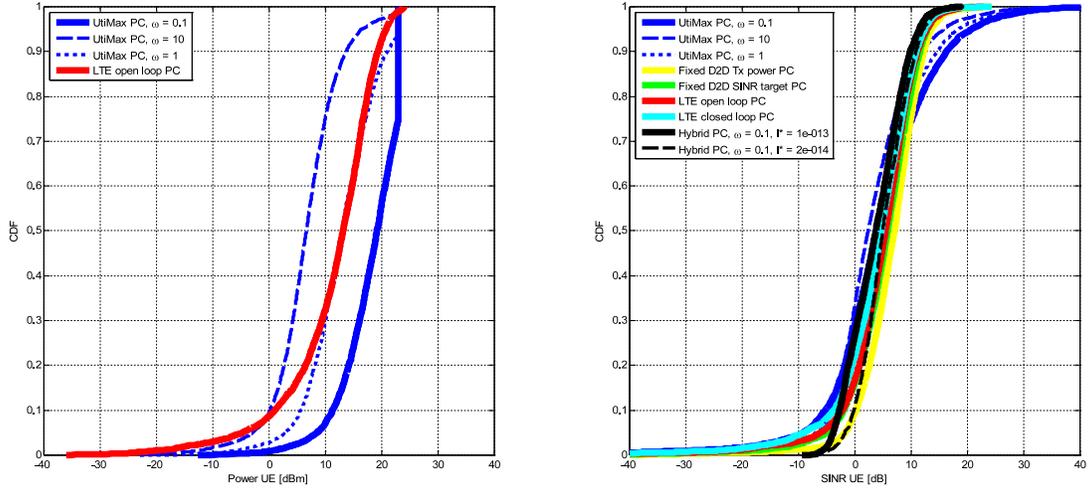}
\vspace{2mm}
\caption{
Comparing the performance of the LTE based open loop fractional path loss compensation
power control with that of the utility optimizing scheme in terms of transmit power and
achieved SINR for the {\it cellular} users.
The utility based scheme can be tuned to boost the SINR performance of the
cellular UEs in the high SINR regime at the expense of higher power consumption by setting
the $\omega$ to a lower value.
For cellular UEs, the hybrid LTE-utility maximizing scheme performs similarly to the LTE
open loop PC.
}
\label{Fig:cellular}
\end{center}
\end{figure}

Figure \ref{Fig:cellular} compares the power consumption and the achieved
SINR of the cellular UEs when employing different PC strategies in the
system.
The cellular UE power consumption is only affected by the PC algorithm
used by the cellular UE (utility maximization or LTE open loop),
as shown by the left hand side figure.
We can see that setting $\omega$ to 1 results in similar power levels for
the utility and LTE based PC schemes, while setting $\omega$ to 0.01 significantly
increases the transmit power level. On the other hand, $\omega=10$ leads to
significant power saving for the cellular UEs as compared to the LTE PC scheme.
The achieved SINR by the cellular UEs depends not only on their own power
control scheme, but also on the power used by the D2D pairs, as is shown by the
right hand side figure.
We can see that in the utility maximization case, setting $\omega$ to a low value
(e.g. when $\omega \leq 1$)
can significantly boost the achieved peak SINR values.
Apart from this high SINR regime, the SINR performance of the hybrid PC scheme
(i.e.\ LTE open loop for the cellular UEs and utility maximization with a low $I^\star$ cap for the D2D pairs)
shows very good performance, showing for example up to 5-8 dB gains in the lower SINR regime
compared to the pure utility maximization schemes (depending on the setting of the $\omega$).
This result shows that setting the $I^\star$ interference limit to a proper value is an
important tool for protecting the cellular traffic from the interference caused by the D2D layer.

\begin{figure}[t!]
\begin{center}
\includegraphics[width=0.85\hsize]{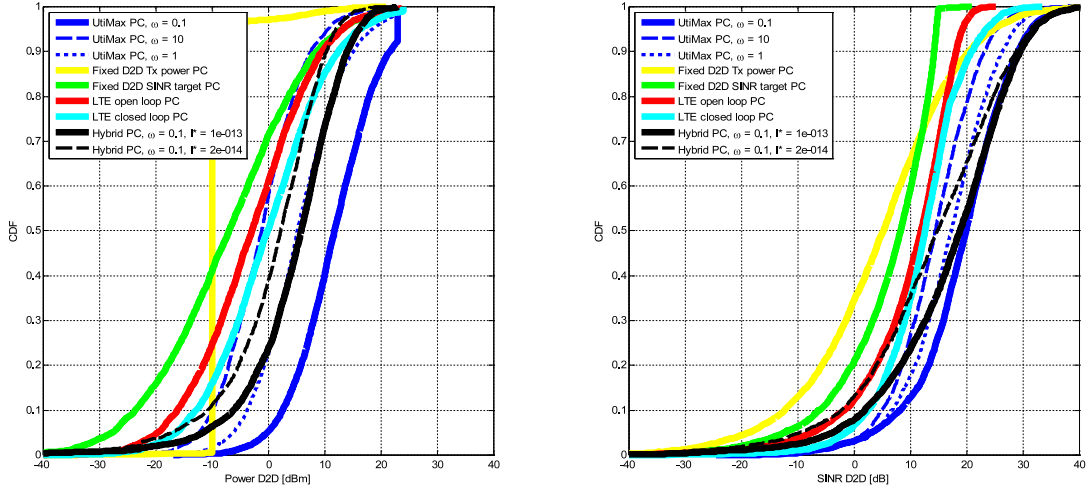}
\caption{
Comparing the transmit power and achieved SINR levels for D2D pairs before mode selection.
With the hybrid scheme, the D2D pairs
enjoy significantly higher SINR values than with legacy LTE-based PC schemes,
at the expense of higher power consumption (depending on the setting of $\omega$).
}
\label{Fig:d2dcand}
\end{center}
\end{figure}

Figure \ref{Fig:d2dcand} shows the distribution of the transmit power and
SINR levels of the D2D pairs.
Similarly to the cellular UEs, the D2D transmit power levels can be tuned
by setting the $\omega$ (here within the range of $\omega=0.01...10$).
We can also see that the different LTE based schemes perform quite differently both
in terms of power consumption and achieved SINR.
In terms of SINR, the LTE open loop power control yields an acceptable performance
(close to the LTE closed loop scheme except in the low SINR region), but this SINR
performance can be significantly improved by employing the hybrid scheme when setting
$\omega$ and $I^\star$ to proper values (e.g. $\omega=1, I^\star=500 \cdot N_0$).
Recall from the previous figure, that the performance punishment for the cellular UEs
when using this more aggressive setting for the D2D pairs is negligible.
When $I^\star$ is set to a low value, the LTE PC scheme performs better in the low and
medium SINR regime.
In Figure \ref{Fig:d2dcand} it is interesting to observe the distribution of the
transmit power level when using the ``Fixed Tx Power'' method for the D2D pairs.
Since this figure shows that transmit power distribution {\it before} mode
selection, the actual power level set for the D2D candidates can be different
from the predetermined fixed transmit power level if the cellular mode is selected
for a D2D candidate.
This is because when using cellular mode, the transmit power level is set by the
open loop path loss compensation method.

%

\vspace{2mm}
\begin{figure}[ht!]
\begin{center}
\includegraphics[width=0.85\hsize]{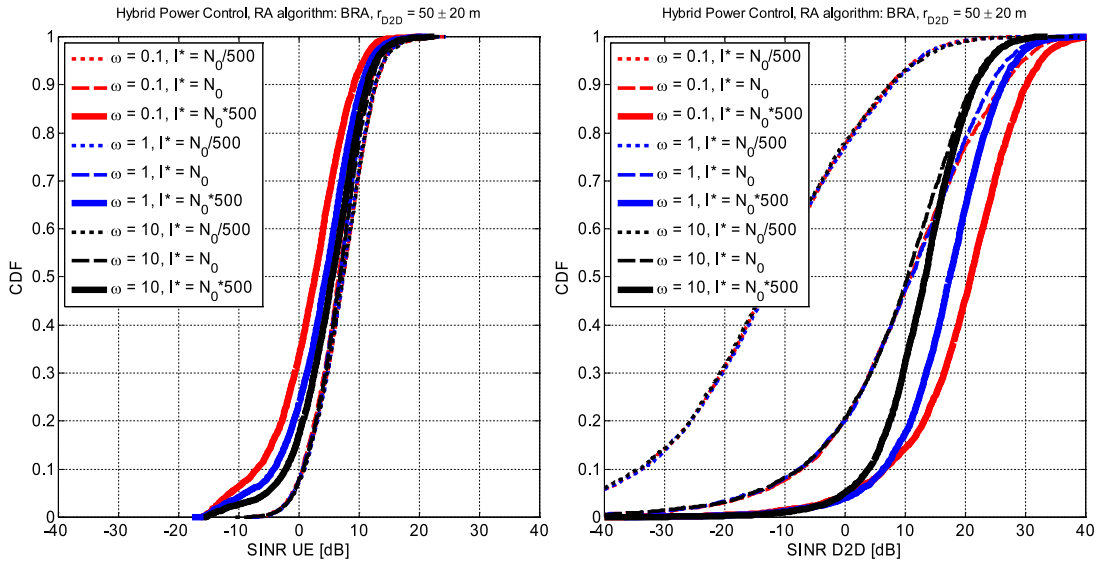}
\caption{
Hybrid PC scheme: cellular UEs use the legacy LTE power control, while D2D users use
the distributed utility maximizing scheme. By setting the $I^\star$ threshold to some suitable value,
(e.g.$\omega=1$, $I^\star=500 \cdot N_0$) the D2D performance can be boosted at the expense of a minimal
impact on the cellular layer.
}
\label{Fig:Hybrid}
\end{center}
\end{figure}

Figure \ref{Fig:Hybrid} examines the trade off between the performances of the cellular and the D2D layers
when using the hybrid PC scheme under
various settings.
Here we can see that setting $\omega=1$ and $I^\star$ to a high value with
respect to the noise floor ($sigma$) boosts the D2D performance (solid blue line) with some moderate and acceptable
negative impact on the cellular layer.

\begin{figure}[t!]
\begin{center}
\includegraphics[width=0.80\hsize]{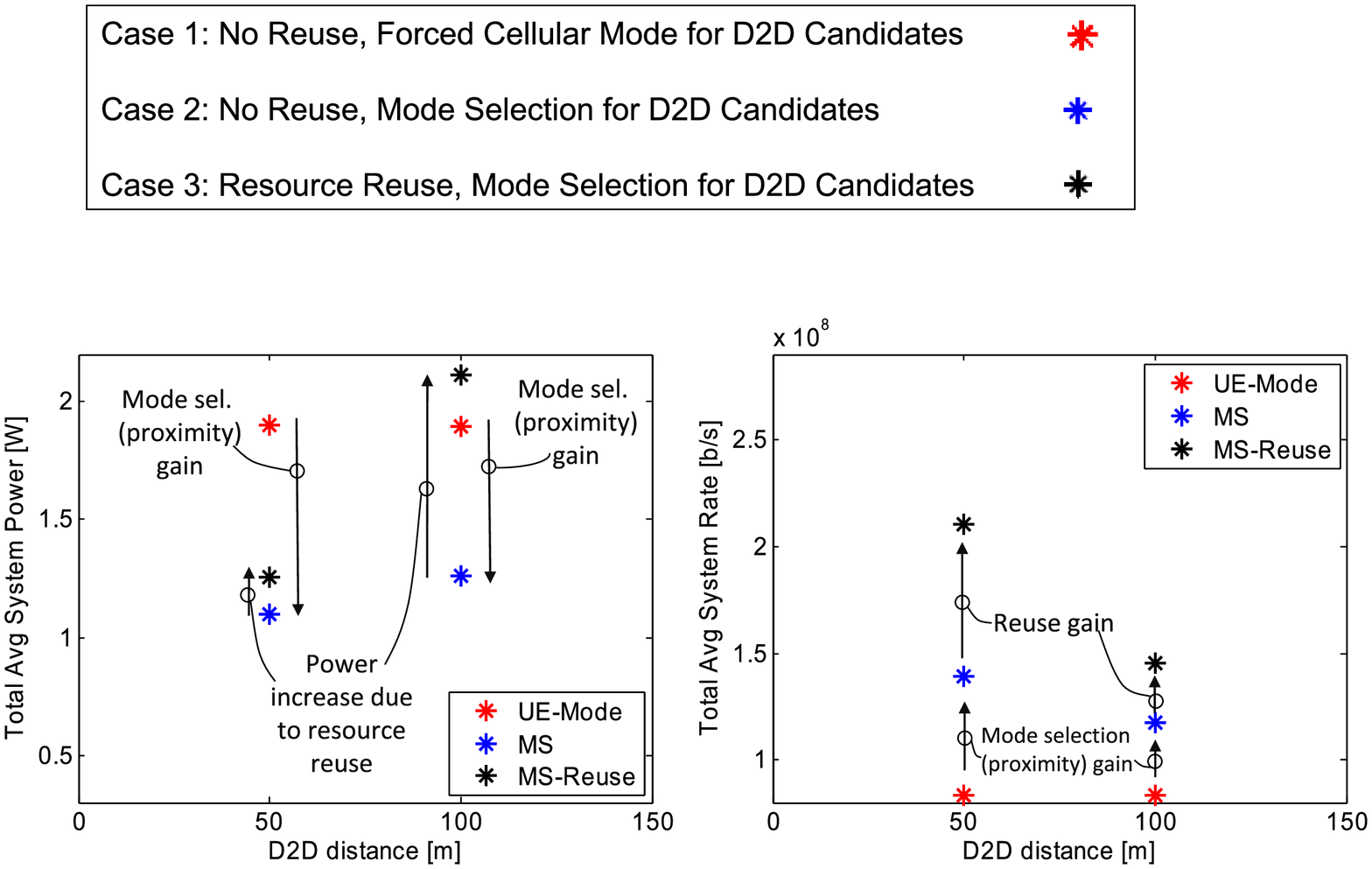}
\caption{
Comparing the mode selection gain and the resource reuse gain in terms of total system throughput and
used power when both cellular UEs and D2D pairs use the LTE open loop PC scheme.
When the D2D candidates are forced to operate in cellular mode (``UE Mode''),
they can neither take advantage
of the short communication distance nor the reuse gain.
When the D2D candidates can select D2D (direct)
mode, proper mode selection (MS) yields increased total system rate and reduced power.
Finally, when D2D pairs
may reuse cellular resources (i.e. a resource block can be used by multiple transmitters), the total
system rate can be further improved (``MS Reuse'').
However,
this total system rate improvement, that is the reuse gain, may come at the expense of some increase of the total system power
(see the left hand side plot, especially when the D2D distance is set to 100m).
}
\label{Fig:gains}
\end{center}
\end{figure}

\begin{figure}[t!]
\begin{center}
\includegraphics[width=0.80\hsize]{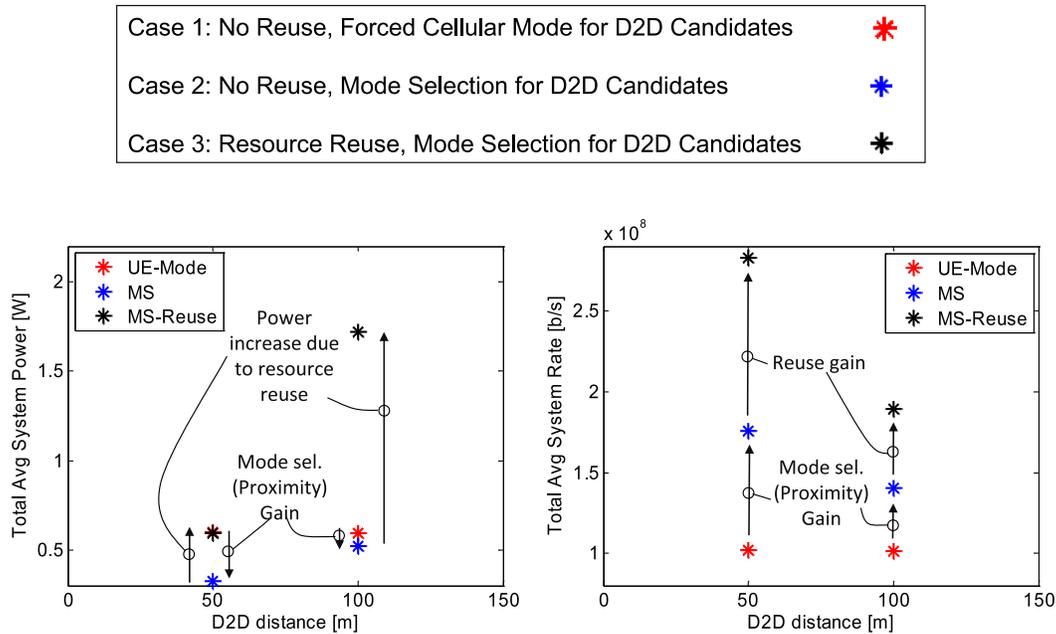}
\caption{
Comparing the mode selection gain and the resource reuse gain in terms of total system throughput and
used power when both the cellular UEs and the D2D pairs use the utility maximization scheme.
The overall gain of the utility optimal scheme (as compared with the LTE based PC ) is quite large both in terms of overall power consumption
and achieved user bit rates, as visible by comparing this figure with Figure \ref{Fig:gains}.
}
\label{Fig:gains2}
\end{center}
\end{figure}

Figures \ref{Fig:gains}-\ref{Fig:gains2} offer an insight into the mode selection
and reuse gains of D2D communications.
Recall that the mode selection gain is due to selecting the direct
communication link rather than using cellular transmission, as is shown in the figure.
When, in addition, resource sharing is possible between D2D and cellular transmitters,
the overall system throughput further increases at a cost of higher total
transmit power.
This total transmit power increase depends on the geometry of the system,
that is exemplified by the right hand side of the figure (i.e.\ total system rate).
We can also observe that the utility maximization significantly improves the
total system rate performance and at same time reducing the average power level
in the system.
(The hybrid scheme (not shown here) performs close to the utility maximization
scheme when $\omega$ and $I^\star$ are properly set.)
%

Finally, Figure \ref{Fig:scatter} shows the correlation between the used
transmit power and achieved SINR levels for cellular UEs and D2D pairs
when using the utility based, the LTE based and the hybrid PC algorithms.
When the LTE PC targeting a fixed SNR level is employed,
the resulting SINR levels are rather similar throughout the simulations.
For the D2D pairs, the fixed Tx power yields a large variation
in the achieved SINR values.
The other LTE based schemes as well as
the utility function method perform in between these two extremes, the
utility based PC providing the best performance in terms of achieved
SINR but tending to consume somewhat higher power both for cellular UEs
and D2D transmitters.

\begin{figure}[t!]
\begin{center}
\caption{
Comparing the transmit power and the SINR performance for cellular UEs (left) and D2D pairs (right)
when D2D pairs employ the LTE and utility PC based algorithms.
For cellular UEs, the utility based
scheme (applied to the D2D pairs) with $\omega=0.1$ tends to trigger higher power values
(set by the LTE OFPC power control) and thereby to reach higher SINR values.
For D2D pairs, the utility based scheme is clearly superior to legacy LTE PC
when comparing the achieved SINR with the same transmit power levels. The hybrid scheme
clearly pushes the D2D performance towards higher SINR values, while essentially keeping
the benefits of the utility maximizing scheme for the cellular UEs, without requiring
a new PC scheme for the cellular UEs.
}
\label{Fig:scatter}
\end{center}
\end{figure}

\section{Conclusions}
\label{Sec:Conc}
In this work, we examined the performance of practical radio resource management algorithms
for D2D communication integrated in cellular networks. The main motivation for this examination
is to gain an understanding of how well LTE friendly power control and resource allocation
schemes perform as compared to optimization based approaches. Specifically, we developed a
distributed power control algorithm that maximizes a utility function that is capable of
balancing between maximizing spectral efficiency and minimizing the sum transmit power for
a given set of interfering D2D and cellular links. We used this algorithm as a benchmarking
tool with respect to practical PC schemes based on the LTE PC toolkit, including ``no power
control'', PC with fixed SINR target, open loop fractional path loss compensation and closed
loop PC. For mode selection and resource allocation, we developed a heuristic algorithm (MinInterf)
that attempts to reduce the intra-cell interference introduced by D2D communications assuming
full path loss knowledge. Using MinInterf as a benchmark, we then examined the performance of
two practically feasible MS and RA algorithms in a realistic system simulator.

The numerical results
indicate that the LTE PC gets close to the utility-based scheme, both in terms of used
transmit power levels by the cellular as well as the D2D users and the resulting SINR values.
The only significant gain with the optimization-based approach is the SINR obtained by the
high performing D2D users.
On the other hand, the LTE OFPC scheme, (depending on the $\omega$ parameter
of the utility-based method) can produce somewhat higher
SINR values for the cellular UEs.
These results tend to suggest that the flexible LTE
power control scheme is well prepared for network assisted D2D communications, especially
for the cellular UEs.
However, for the D2D pairs, the utility based scheme can provide gains in terms of
SINR distribution and total transmit power consumption.
These gains can be harvested by a hybrid scheme, in which cellular UEs use the LTE PC scheme,
whereas D2D users rely on a distributed scheme, whose parameters in practice can be controlled
by the cellular network. In future work we plan to investigate methods to set the value of $I^\star$.
%

\section*{Appendix A: Derivation of Inequality \eqref{inequality}}
Constraints in Problem \eqref{problem7} can be elaborated by appealing to the definition of matrix $\mathbf{H}^T$ in Equation \eqref{eq:H} as follows:
\begin{eqnarray}
\mathbf{H}^T \boldsymbol{\lambda}^{\text{\tiny(LP)}} &\succeq& - \omega \mathbf{1} \nonumber \\
\sum_k h_{kl} \lambda_k^{\text{\tiny(LP)}} &\geq& - \omega, \quad \forall l; \nonumber \\
-h_{ll} \lambda_l^{\text{\tiny(LP)}} + \displaystyle \sum_{k\neq l} h_{kl} \lambda_k^{\text{\tiny(LP)}} &\geq& - \omega,  \quad \forall l \nonumber \\
\frac{ \lambda_l^{\text{\tiny(LP)}} }{\omega} - \displaystyle \sum_{k\neq l} \frac{G_{kl}}{G_{kk}} \gamma_k^{tgt} \frac{ \lambda_k^{\text{\tiny(LP)}}}{\omega} &\leq& 1 ,\quad \forall l. \nonumber \hspace{1.5cm} \blacksquare
\end{eqnarray}
\section*{Appendix B: Derivation of Inequality \eqref{SIR}}
Inequality $\eqref{SIR}$ can be derived by appealing to Equation \eqref{mu} as follows:
\vspace{-1mm}
\begin{eqnarray}
\frac{\lambda_l^{\text{\tiny(LP)}}}{\omega} - \displaystyle \sum_{k\neq l} \frac{G_{kl}}{G_{kk}} \gamma_k^{tgt} \frac{\lambda_k^{\text{\tiny(LP)}}} {\omega} &\leq& 1 \nonumber \\
\frac{\lambda_l^{\text{\tiny(LP)}}}{\omega} \frac{\gamma_l^{tgt} \sigma_l}{G_{ll}}  \frac{G_{ll}}{\gamma_l^{tgt} \sigma_l } - \displaystyle \sum_{k\neq l} \frac{G_{kl}}{G_{kk}} \gamma_k^{tgt} \frac{ \lambda_k^{\text{\tiny(LP)}}}{\omega} \frac{\sigma_k}{\sigma_k} &\leq& 1 \nonumber \\
\frac{\mu_l G_{ll}}{\gamma_l^{tgt} \sigma_l} &\leq& \displaystyle \sum_{k\neq l} \frac{G_{kl}}{\sigma_k}\mu_k +1 \nonumber \\
\gamma_l^{cc} (\boldsymbol{\mu}) \triangleq \frac{\mu_l G_{ll}}{\sigma_l + \displaystyle \sum_{k\neq l} G_{kl} \frac{\sigma_l}{\sigma_k}\mu_k} &\leq& \gamma_l^{tgt}, \quad \forall l. \nonumber \hspace{2cm} \blacksquare
\end{eqnarray}
\bibliography{D2D_journal}
\end{document}